\documentclass[aps,prb,twocolumn,amsmath,amssymb,superscriptaddress,notitlepage]{revtex4-2}
\usepackage{graphicx}
\usepackage{amsmath,amssymb}
\usepackage{dcolumn} 
\usepackage{bm} 
\usepackage{xcolor}
\usepackage[utf8]{inputenc}
\usepackage{appendix}
\usepackage{listings}
\usepackage{float}
\usepackage{epstopdf}
\usepackage{hyperref}
\usepackage[all]{hypcap}
\usepackage{geometry}
\usepackage{natbib}
\usepackage{soul}

\begin{document}

\title{Non-local thermal transport modeling using the thermal distributor}
\author{Ali Kefayati}
\email{alikefay@buffalo.edu}
\affiliation{Department of Electrical Engineering, University at Buffalo, The State University of New York, Buffalo, New York 14260, USA.}
\author{Philip B. Allen}
\email{philip.allen@stonybrook.edu}
\affiliation{Department of Physics and Astronomy, Stony Brook University, Stony Brook, New York 11794-3800, USA.}
\author{Vasili Perebeinos}
\email{vasilipe@buffalo.edu}
\affiliation{Department of Electrical Engineering, University at Buffalo, The State University of New York, Buffalo, New York 14260, USA.}

\date{\today}

\begin{abstract}

Thermal transport in a quasi-ballistic regime is determined not only by the local temperature  $T(r)$, or its gradient  $\nabla T(r)$, but also by temperature distribution at neighboring points. For an accurate description of non-local effects on thermal transport, we employ the thermal distributor, $\Theta (r,r')$, which provides the temperature response of the system at point $r$ to the heat input at point $r'$. We determine the thermal distributors from the linearized Peierls-Boltzmann equation (LPBE), 
both with and without the relaxation time approximation (RTA), and employ them to describe thermal transport in quasi-ballistic graphene devices. 
\end{abstract}

\maketitle

\section{\label{sec:level1}Introduction}
 
 Advances in technology and experimental studies beyond micro-scale dimensions of materials 
 require new insights into theoretical models that had been developed initially based on the continuum 
 transport theories. The Peierls formulation of thermal transport in solids
(the Peierls-Boltzmann equation, PBE \cite{Peierls29}) is based on the
 quasiparticle picture of phonons.  The temperature gradient, $\nabla T$, enters the PBE as a driving force. 
 At macroscopic scales and in steady-state, the PBE leads to the Fourier's law, $\vec{J}=-\kappa(T_0)\nabla T$,
 where $\kappa(T_0)$ is the thermal conductivity at the background temperature, $T_0$, and $\vec{J}$ is the heat current 
 density.  Small variations of the temperature gradient are ignored. This is called the diffusive regime.

Early experiments on thermal transport in submicron devices
\cite{simons1960boltzmann,levinson1980nonlocal,mahan1988nonlocal,majumdar1993microscale,chen2005nanoscale} 
showed that the temperature gradient is not constant, but varies on length scales shorter than or comparable to 
the mean free path (MFP) of phonons.  Often analysis of experiment suggests a version of
Fourier's law using an ``effective thermal conductivity.''
The experiments indicate a non-diffusive regime, with a non-local relation between heat current 
and temperature gradient. Fourier’s law requires generalization, and Boltzmann theory does this well ~\cite{mahan1988nonlocal,ordonez2011constitutive,allen2018temperature, hua2019generalized, hua2020space,
simoncelli2020generalization}.
The Boltzmann equation describes the evolution of the phonon distribution function
$N_Q(r,t)$. When phonons are driven away from equilibrium by local power insertion, it is necessary to add a new term $(dN_Q/dt)_{\rm ext}$ describing the power $P_Q(r,t)$ added to phonon mode $Q$. The local temperature $T(r,t)=T_0+\Delta T(r,t)$ is an ultimate goal, but is not needed to find the non-equilibrium distribution. The inserted power $P_Q(r,t)$ is enough to determine $N_Q$ and the corresponding heat current $\vec{J}(r,t)$. The local temperature deviation $\Delta T(r,t)$ is an important measure of the behavior of the system, but Boltzmann theory does not contain a definition of $\Delta T(r,t)$; it is necessary to choose a definition. The correct definition is that $C\Delta T(r,t)=\Delta E(r,t)$, where $\Delta E(r,t)$ is the deviation of the total non-equilibrium phonon energy of the system when it is driven away from the equilibrium state at the background temperature $T_0$, and $C$ is the specific heat. Unfortunately, when the Boltzmann scattering operator is approximated by its relaxation time approximation (or RTA), an alternate and less physical definition is necessary to restore the energy conservation that is broken by RTA. In this paper, we find $\Delta T(r,t)$ by solving the linearized PBE (or LPBE) using the full scattering operator and the correct definition, and compare it with the RTA version.

Solving the PBE requires a matrix inversion, which is often avoided by using the relaxation time approximation,
\begin{equation}
\left(\frac{\partial N_Q}{\partial t}\right)_{\rm scatt}^{\rm RTA}=-\frac{N_Q(\vec{r},t)-n_Q(T(\vec{r},t))}{\tau_Q}.
\label{eq:}
\end{equation}
Here $Q=(\vec{q},s)$ labels phonon modes: $\vec{q}$ is the wavevector, and $s$ is the branch index. The Bose-Einstein distribution $n_Q(T(\vec{r},t))$ is evaluated at the local temperature $T(\vec{r},t)$. The phonon relaxation rate $1/\tau_Q$ is evaluated using the Fermi golden rule for anharmonic three-
phonon scatterings. It is also the diagonal part of the linearized scattering operator, $\hat{S}^0$,
\begin{equation}
\left(\frac{\partial N_Q}{\partial t}\right)_{\rm scatt}^{\rm LPBE}=-\sum_{Q^\prime} S_{QQ^\prime}^0 (N_{Q^\prime}-n_{Q^\prime}).
\label{eq:}
\end{equation}
In this version labeled with superscript 0, $1/\tau_Q=S_{QQ}^0$.   The correctly linearized operator 
$\hat{S}^0$ is non-Hermitian.  For numerical inversion, it is preferable instead to define
\cite{srivastava2019physics},
\begin{eqnarray}
N_Q(\vec{r},t)&\equiv& n_Q(T(\vec{r},t))+
\nonumber \\
&+&n_Q^0(T_0)(n_Q^0(T_0)+1)\phi_Q(\vec{r},t)
\label{Equ:NQ}
\end{eqnarray}
\begin{equation}
\left(\frac{\partial N_Q}{\partial t}\right)_{\rm scatt}^{\rm LPBE}=-\sum_{Q^\prime} S_{QQ^\prime} \phi_{Q^\prime}.
\label{Equ:Scatt}
\end{equation}
where the $n_Q^0(T_0)$ is the Bose-Einstein distribution at equilibrium temperature $T_0$. Then the operator $\hat{S}$ is Hermitian, and the diagonal element is
\begin{equation}
S_{QQ}=n_Q^0(T_0)[n_Q^0(T_0)+1]/\tau_Q.
\label{eq:Sdiag}
\end{equation}
%

 
The spatially homogeneous PBE driven by a constant $\nabla T$ has been solved
by inversion of this Hermitian operator $\hat{S}$ \cite{lindsay2014phonon, lindsay2016first, mcgaughey2019phonon,
cepellotti2016thermal, cepellotti2017transport, fugallo2013ab, fugallo2014thermal, li2014shengbte, 
CHERNATYNSKIY2015196, carrete2017almabte}.  For spatially inhomogeneous situations, 
the LPBE (in Fourier space $(\vec{k},\eta)$ rather than
coordinate space $(\vec{r},t)$) requires much more difficult inversion of the non-Hermitian
operator $S_{QQ^\prime}+i(\vec{k}\cdot\vec{v}_Q-\eta)\delta_{QQ^\prime}$ 
where $\vec{v}_Q$ is the velocity of the phonon mode $Q$.  The difficult inversion is avoided by using the RTA approximation $S_{QQ^\prime}\rightarrow \delta_{QQ^\prime}n_Q(n_Q+1)/\tau_Q$
\cite{hua2014analytical, zeng2014disparate, collins2013non, allen2018analysis}.  Recently 
inversions with the correct scattering operator for inhomogeneous transport have been done
 \cite{hua2020space,chiloyan2021green}.  

\begin{figure}
  \centering
 \includegraphics[width=0.5\textwidth]{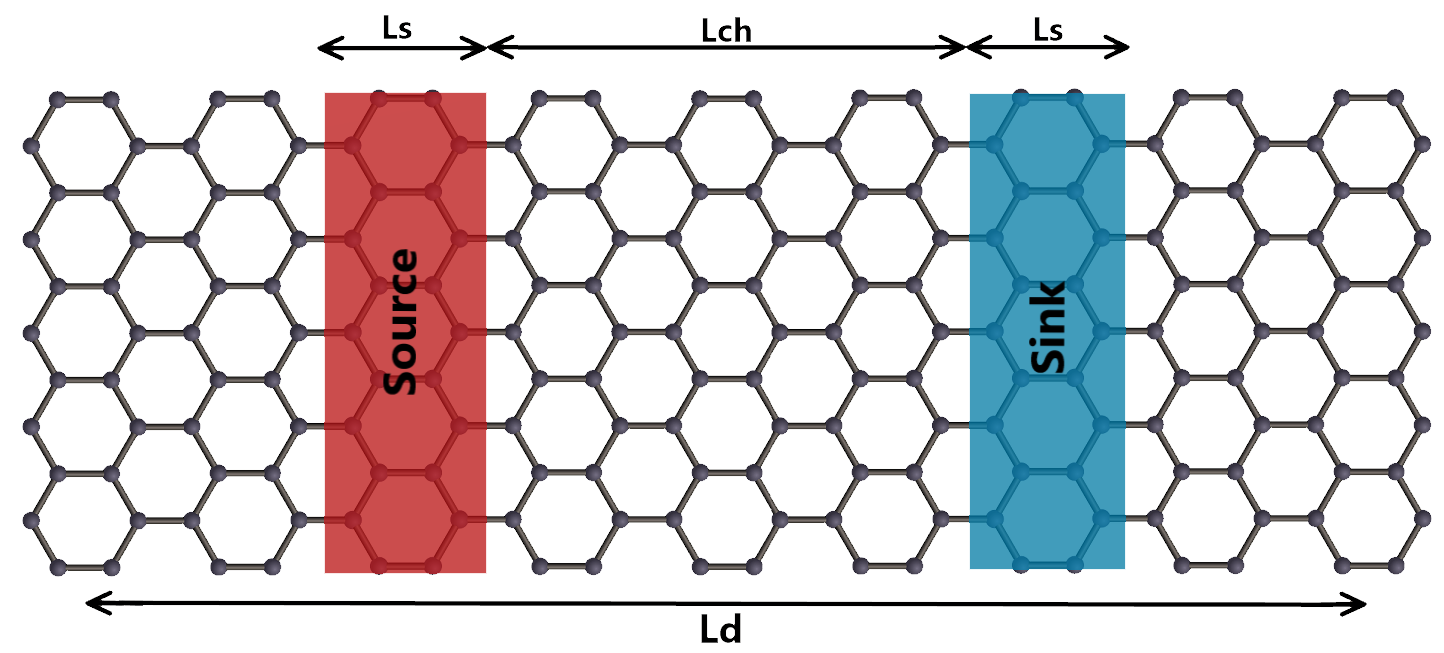}
 \caption{Schematic of inhomogeneous external driving with periodic boundary conditions.
 The finite system has a length $L_{\rm d}=2L_{\rm s}+2L_{\rm ch}$, which is repeated periodically, where $L_{\rm ch}$ is the channel length and $L_{\rm s}$ is the source/sink heat length.
 Thermal energy at rate $P$ is added at the source and removed at the sink.  }
\label{fig:graphenemodel}
\end{figure}

In \cite{allen2018temperature}, the authors introduced a new concept called thermal susceptibility, inspired by
the definition of the electrical susceptibility. Thermal susceptibility relates the temperature deviation at $(\vec{r},t)$ 
to the heat insertion at $(\vec{r}^{ \ \prime},t^{\prime})$. This study aims to investigate the capability of the \textit{thermal distributor} function $\Theta$, 
which is a redefined version of the thermal susceptibility function:
\begin{equation}
\Theta(\vec{r}-\vec{r}^{ \ \prime}, t-t^{\prime})\equiv \frac{\delta T(\vec{r},t)}{\delta P(\vec{r}^{ \ \prime},t^{\prime})}
\label{eq:deftheta}
\end{equation}
for the analysis of non-local thermal transport. Thus the temperature deviation is obtained, 
\begin{equation}
\resizebox{.98\hsize}{!}{$\Delta T(\vec{r},t) = \frac{1}{V}\int d\vec{r}^{ \ \prime} \int_{-\infty}^t dt^\prime
 \Theta(\vec{r}-\vec{r}^{ \ \prime},t-t^{\prime}) P(\vec{r}^{ \ \prime},t^{\prime}),
\label{eq:defdT}$}
\end{equation}
where $V$ is the sample volume. 
In reciprocal space, $\Delta T(\vec{k},\eta)=\Theta(\vec{k},\eta)P(\vec{k},\eta)$.
We apply our analysis to graphene, depicted in Fig. \ref{fig:graphenemodel}.

Because graphene is a two-dimensional crystal, the vectors $\vec{r}$ and $\vec{k}$ are two-dimensional.
Because the heat source and sink are parallel to the $y$ axis, the relevant wavevector is $\vec{k}=(k_x,0)$.
Heat current density $\vec{J}_{\rm 2D}$ has units W/m; the more familiar unit in 3D is W/m$^2$.  Thermal conductivity
in 2D has units W/K; in order to compare with 3D, it is conventional to choose the somewhat arbitrary 
``thickness'' for graphene to be $h=3.4$ \AA. 
 This paper uses 3D units for current density $J=J_{\rm 2D}/h$, input power $P=P_{\rm 2D}/h$, 
 energy density $U=U_{\rm 2D}/h$, and specific heat $C=C_{\rm 2D}/h$.  Then $\kappa$ has
 conventional units W/mK, and $\Theta$ has units Km$^3$/W.

The measured thermal conductivity of graphene (in 3D units) is reported to lie in the range of 2600 to 5300 W/mK 
at room temperature~\cite{balandin2008superior, chen2011raman}. The theoretical thermal conductivity of 
pristine infinite-size graphene at room temperature is 
reported in the range of 2800-4300 W/mK~\cite{lindsay2014phonon, fugallo2014thermal, libbi2020thermomechanical}. 
In devices smaller than mean free paths $\Lambda_Q$ of important phonons, the measured heat current divided by
an approximate measurement of temperature gradient gives an ``effective thermal conductivity'' ($\kappa_{\rm eff}$)
of smaller value.  For phonons with bulk $\Lambda_Q=|v_{Qx}|\tau_Q$ ($\vec{v}_Q$ is phonon group velocity)
greater than device size $L$, the contribution to
$\kappa_{\rm eff}$ is reduced from $C_Q|v_{Q,x}|\Lambda_Q $ to $C_Q |v_{Qx}| L$,
where $C_Q$ is the contribution of mode $Q$ to the specific heat.  We will describe this effect using the thermal distributor function \cite{hua2014analytical}.


\section{\label{sec:level1}Formalism}

Under the assumptions of well-defined quasiparticles, the PBE in a crystalline solid is:
\begin{equation}
    \frac{dN_Q}{dt} = \left( \frac{\partial N_Q}{\partial t} \right)_{\rm drift} + \left( \frac{\partial N_Q}{\partial t} \right)_{\rm scat} + 
    \left( \frac{\partial N_Q}{\partial t} \right)_{\rm ext},
    \label{Equ:PBE}
\end{equation}
The first term on the right-hand side of Eq. \ref{Equ:PBE} is the change of $N_Q$ caused by 
phonon drift in the distribution gradient: 
\begin{equation}
    \left( \frac{\partial N_Q}{\partial t} \right)_{\rm drift} = -\vec{v_Q}.\vec{\nabla}_{\vec{r}} N_Q
\label{Equ:drift}
\end{equation}
where $\vec{\nabla}_{\vec{r}}$ is the spatial gradient. 
The second term in Eq. \ref{Equ:PBE} contains all scattering processes in the crystal.  The term from anharmonic
three-phonon scatterings ($Q\rightarrow Q'+Q'', Q+Q'\rightarrow Q''$)  can be found from the Fermi golden rule~\cite{Bonini2012}:
\begin{widetext}
\begin{equation}
\begin{split}
        \left( \frac{\partial N_Q}{\partial t} \right)_{\rm scat} &= \frac{\pi\hbar}{16N_q} \sum_{Q'Q''}\left| V_{QQ'Q''} \right|^2 \\ &\frac{1}{2} 
        \bigg\{ N_Q(N_{Q'}+1)(N_{Q''}+1)-(N_Q+1)N_{Q'}N_{Q''} \bigg\} \delta(\omega_Q-\omega_{Q'} -\omega_{Q''}) \\
        &+ \bigg\{ N_QN_{Q'}(N_{Q''}+1)-(N_Q+1)(N_{Q'}+1)N_{Q''} \bigg\} \delta(\omega_Q+\omega_{Q'} -\omega_{Q''}),
\end{split}
\label{Equ:FGR}
\end{equation}
\end{widetext}
where $\omega_Q$ is the phonon frequency and $N_q$ is a number of wavevectors in the Brillouin zone~\footnote{In the simulations, we use $N_q=120\times120$ $q$-points to sample 
the Brillouin zone of graphene and for delta functions in Eq.~(\ref{Equ:FGR}) we use Gaussian broadening function $\delta(x)=\exp{-(x/\sigma)^2}/\sqrt{\pi}\sigma$ with $\hbar\sigma=0.9$ meV.}.  
$V_{QQ'Q''}$ is the matrix element of the three-phonon process, given by,
\begin{widetext}
\begin{equation}
\begin{split}
         V_{QQ'Q''} = \sum_{MLnml}\sum_{\alpha\beta\gamma}\frac{\epsilon_{Q}^{n\alpha}\epsilon_{Q'}^{m\beta}\epsilon_{Q''}^{l\gamma}}
         {\sqrt{M_{\rm c}^3\omega_Q\omega_{Q'}\omega_{Q''}}} \Psi_{\alpha\beta\gamma}(0n,Mm,Ll)e^{i\boldsymbol{q'.R_M}}
         e^{i\boldsymbol{q''.R_L}}\delta(q+q'+q'',G),
\end{split}
\label{Equ:ME}
\end{equation}
\end{widetext}
where $\Psi_{\alpha\beta\gamma}(0n,Mm,Ll)$ is the third derivative of the crystal potential by displacements 
of atoms in positions $(n,m,l)$ inside the unit cells $(0,M,L)$.  The supercell with index $0$ is the central unit cell. 
$\epsilon_{Q}^{n\alpha}$ is the $\alpha^{\rm th}$ Cartesian component of the polarization vector of mode $Q$ 
at atom $n$, and $M_{\rm c}$ is the mass of a carbon atom.  The Kronecker delta ensures the conservation of the lattice momentum, where $G$ is a reciprocal lattice vector. 
The local equilibrium phonon population $n_Q=n_Q(T(\vec{r},t))$ depends implicitly on the space and time through its explicit dependence 
on the temperature $T(\vec{r},t)$.  For small deviations from equilibrium, expand the  
phonon population $N_Q$ as in Eq. \ref{Equ:NQ} to first order in $\phi_Q$.
The anharmonic scattering matrix $\hat{S}$ (Eq. \ref{Equ:Scatt}) then has diagonal and off-diagonal elements,
\begin{widetext}
\begin{eqnarray}
        \nonumber 
        S_{QQ} &=&1/\tau_Q= 2\pi\hbar \sum_{Q'Q''}\left| V_{QQ'Q''} \right|^2 
        \\ \nonumber && \bigg\{ \frac{(n_Q^0+1)n_{Q'}^0n_{Q''}^0}
        {2}\delta(\omega_Q-\omega_{Q'} -\omega_{Q''}) +n_Q^0n_{Q'}^0(n_{Q''}^0+1)
        \delta(\omega_Q+\omega_{Q'} -\omega_{Q''}) \bigg\}  
        \\ \nonumber 
        S_{QQ'} &=& 2\pi\hbar \sum_{Q''}\left| V_{QQ'Q''} \right|^2 \bigg\{ n_Q^0n_{Q'}^0(n_{Q''}^0+1)
        \delta(\omega_Q+\omega_{Q'} -\omega_{Q''}) \\  &&  -(n_Q^0+1)n_{Q'}^0n_{Q''}^0
        \delta(\omega_Q-\omega_{Q'} -\omega_{Q''}) -n_Q^0(n_{Q'}^0+1)n_{Q''}^0
        \delta(\omega_Q-\omega_{Q'} +\omega_{Q''}) \bigg\}
\label{Equ:SE}
\end{eqnarray}
\end{widetext}
This version of the scattering matrix is real-symmetric ($S_{QQ'}=S_{Q'Q}$) {\it i.e.} Hermitian. Each collision conserves
phonon energy, which is assured by 
\begin{equation}
S_{QQ}\omega_{Q}+\sum_{Q', Q'\ne Q}S_{QQ'}\omega_{Q'}=0, \ {\rm or} \ \hat{S}|\omega\rangle=0.
\label{eq:}
\end{equation}
The mode frequency $\omega_Q=\langle Q|\omega\rangle$ is an eigenvector,
in fact, the {\bf only} ``null eigenvector'', of the linearized
Hermitian scattering operator $S_{QQ^\prime}=\langle Q|\hat{S}|Q^\prime\rangle$.

The last term in Eq.~(\ref{Equ:PBE})  models external heat sources and sinks.
The form is usually \cite{hua2018heat,chiloyan2020thermal}
\begin{equation}
    \left( \frac{\partial N_Q}{\partial t} \right)_{\rm ext} = \frac{P_Q(\vec{r},t)}{C}\frac{dn_Q}{dT},
\label{Equ:ext}
\end{equation}
The heat source/sink $P$, its geometry $(\vec{r},t)$, and its spectral distribution $Q$ determine whether 
the heat transport is quasiballistic or diffusive.  We use the simplest version where $P_Q=P$ is independent of $Q$
\cite{Vermeersch2014,VermeerschI2015,allen2018temperature,chiloyan2020thermal}, 
\begin{equation}
    \left( \frac{\partial N_Q}{\partial t} \right)_{\rm ext} = \frac{P}{C}\frac{dn_Q}{dT},
\label{Equ:ext}
\end{equation}
where $P$ is the heat power added per unit volume of the system.  
Each mode gets the same boost $\Delta T$ from $P(\vec{r},t)$.  Detailed knowledge of source and sink would cause a $Q$-dependence of $P$ \cite{hua2014analytical,Allen2022}, but missing this knowledge,
$P_Q=P$ is a sensible guess.

In vector-space notation, the LPBE,  Eq. \ref{Equ:PBE}, is
\begin{widetext}
\begin{equation}
    \frac{\partial}{\partial t} [ |n\rangle+|n^0(n^0+1)\phi\rangle ] = -\vec{v_Q}\vec{\nabla}_{\vec{r}}[ |n\rangle+|n^0(n^0+1)\phi\rangle ] 
    - \hat{S} |\phi\rangle+\frac{P(\vec{r},t)}{C} |\frac{dn}{dT}\rangle.
\label{Equ:PBEVec}
\end{equation}
\end{widetext}
In this notation, the kets (like $|n\rangle$) are vectors in the space of phonon modes,
with components $\langle Q|n\rangle=n_Q$.
The solution $\phi_Q(\vec{r},t)$ is found from its Fourier  $(\vec{k},\eta)$ representation, 
\begin{equation}
    \phi_Q(\vec{r},t) = \frac{1}{2\pi} \int_{-\infty}^{\infty} d\eta e^{-i\eta t}
     \sum_{\vec{k}} e^{i\vec{k}\cdot\vec{r}} \phi (\vec{k},\eta).
\label{Equ:Phi}
\end{equation}
The temperature deviation $\Delta T(\vec{r},t)$
 and the power input $P(\vec{r},t)$ are also transformed to Fourier space.  To simplify the
 algebra, define a vector $|X\rangle$ and an operator $\hat{W}$.
\begin{equation}
    |X\rangle \equiv |n^0(n^0+1)\hbar \omega\rangle
\label{Equ:XQ}
\end{equation}
\begin{equation}
\hat{W} \equiv \hat{S}+ i(\vec{k} \cdot \hat{\vec{v}} - \eta \hat{1})  \hat{n}^0(\hat{n}^0+\hat{1})
\label{Equ:W}
\end{equation}
where $\hat{\vec{v}}$ and $\hat{n}^0$ are diagonal in $Q$ space ({\it i.e.} 
$\langle Q|\hat{n}^0| Q^\prime \rangle = n_Q^0 \delta(Q,Q^\prime)=\langle Q|n^0\rangle \delta(Q,Q^\prime)$).
The LPBE in Fourier space is
%
\begin{equation}
\begin{split}
    \hat{W}|\phi\rangle =  \frac{1}{k_BT^2} &\Big[ -i(\vec{k}\cdot \hat{\vec{v}}-\eta\hat{1})\Delta T(\vec{k},\eta) + \\
    &\frac{P(\vec{k},\eta)}{C} 
\hat{1} \Big] | X\rangle
\end{split}
\label{Equ:PBEWX}
\end{equation}
%

\subsection{Thermal distributor.} 

By inverting the matrix $\hat{W}$, the distribution $|\phi(\vec{k},\eta)\rangle$ is found.   
The non-equilibrium energy density, $\Delta U(\vec{r},t)$, is the total local energy density minus the 
energy density of the system equilibrated at the local temperature $T(\vec{r},t)=T_0 + \Delta T(\vec{r},t)$.  
Its Fourier version is 
 \begin{widetext}
 \begin{equation}
     \Delta U(\vec{k},\eta) = \frac{1}{V} \langle X|\phi \rangle = \frac{1}{Vk_BT^2} \left [ \langle X| \hat{W}^{-1}|
     (-i(\vec{k}\cdot \vec{v}-\eta))X\rangle \Delta T (\vec{k},\eta) + \langle X| \hat{W}^{-1} |X\rangle \frac{P(\vec{k},\eta)}{C}    \right ]
\label{Equ:EC}
\end{equation}
\end{widetext}
After transients have died out, the local equilibrium part $|n(T(\vec{r},t))\rangle$ of the distribution contains
all the heat and the deviation  $|n_0(n_0+1)\phi\rangle$ contains no net heat.  
This is a result of Boltzmann’s H theorem, which says that before a steady state is reached, collisions increase entropy. The steady-state occurs when entropy is maximum, which happens when the distribution evolves to a Bose function $|n(T(\vec{r},t))\rangle$ that contains all the heat energy \cite{allen2018temperature}.  Therefore $\Delta U(\vec{k},\eta) =0$, and the thermal distributor function 
$\Theta(\vec{k},\eta)$, defined in Eq. \ref{eq:defdT}, can be calculated from Eq. \ref{Equ:EC} as:
\begin{equation}
    \Theta(\vec{k},\eta) = \frac{\Delta T(\vec{k},\eta)}{P(\vec{k},\eta)}  = \frac{1}{C}\frac{\langle X| \hat{W}^{-1}|X\rangle}
    {\langle X| \hat{W}^{-1} | i(\vec{k} \cdot \vec{v}-\eta)X\rangle}.
\label{Equ:Theta}
\end{equation}
This is the linear relation that gives the local temperature deviation caused by the external heat power input $P$.

\subsection{Thermal conductivity.}  

The thermal current, using Eq.~(\ref{Equ:PBEWX}), is  
\begin{widetext}
\begin{equation}
\begin{split}
   \vec{J}(\vec{k},\eta)  &= \sum_Q \hbar\omega_Q \vec{v}_Q n_Q^0 (n_Q^0+1)\phi_Q(\vec{k},\omega) \\
   &= \frac{1}{V} \langle \vec{v}X|\phi\rangle = \frac{1}{Vk_BT^2} \left [ \langle \vec{v}X| \hat{W}^{-1} |
   [-i(\vec{k} \cdot \vec{v}-\eta)]X\rangle \Delta T (\vec{k},\eta) + \langle \vec{v}X| \hat{W}^{-1} |X\rangle \frac{P(\vec{k},\eta)}{C}    \right ]
\end{split}
\label{Equ:J}
\end{equation}
\end{widetext}
In Fourier space, the Fourier's law reads: $\vec{J}(\vec{k},\eta)=-i\vec{k}\kappa(\vec{k},\eta)\Delta T(\vec{k},\eta)$, 
and the thermal conductivity according to Eqs. (\ref{Equ:Theta}),(\ref{Equ:J}) is  
\begin{widetext}
\begin{equation}
\begin{split}
    \kappa(\vec{k},\eta) &=\frac{i\vec{k}\vec{J}(\vec{k},\eta)}{k^2\Delta T(\vec{k},\eta)} \\
    &= \frac{1}{k^2Vk_BT^2} 
    \bigg [ 
    \langle i\vec{k}\cdot\vec{v}X| \hat{W} ^{-1}|[-i(\vec{k}\cdot\vec{v}-\eta)]X\rangle 
    + \langle i\vec{k}\cdot\vec{v}X| \hat{W} ^{-1}|X\rangle \frac{\langle X| \hat{W} ^{-1}|i(\vec{k} \cdot\vec{v}-\eta)X\rangle}
    {\langle X| \hat{W} ^{-1}|X\rangle}
  \bigg ].
\end{split}
\label{Equ:Kap}
\end{equation}
\end{widetext}
Using Eq.~(\ref{Equ:XQ}) and (\ref{Equ:W}) we can write 
$|i(\vec{k} \cdot\vec{v}-\eta)X\rangle$ as $( \hat{W} -\hat{S}) | \hbar\omega\rangle$. Using time-reversal symmetry, 
$\hat{S}|\hbar\omega\rangle=0$, and $\langle X|\hbar\omega\rangle=CVk_BT^2$, we can simplify 
Eq. \ref{Equ:Kap} to: 
\begin{equation}
    \kappa(\vec{k},\eta) = \frac{C}{k^2} \left [ \frac{\langle i\vec{k}\vec{v}X| \hat{W} ^{-1}|X\rangle}{\langle X| \hat{W} ^{-1}|X\rangle}  \right ]
    \label{Equ:Appx5}
\end{equation}

By comparing Eq.~(\ref{Equ:Appx5}) with Eq.~(\ref{Equ:Theta}), the relation between the thermal distributor and thermal conductivity is \cite{allen2018temperature}: 
\begin{equation}
    \kappa(\vec{k},\eta) = \frac{1}{k^2}\left(\frac{1}{\Theta(\vec{k},\eta)}+iC\eta\right)
\label{Equ:ThetaKappa}
\end{equation}

Recently it has been shown \cite{hua2019generalized,hua2020space} that unless $P_Q $ is
independent of mode $Q$, the response function $\kappa(\vec{k},\eta)$ is not a full description of non-local
thermal heat transport.  The current in Fourier space has the more general form
 $J(\vec{k},\eta) = -\kappa(\vec{k},\eta) \nabla T(\vec{k},\eta)+B(\vec{k},\eta)$, where $B$ vanishes if
 $P$ is independent of $Q$.  We agree and find that the thermal distributor also needs modification.
 Specifically, the temperature in Fourier space takes the form
$\Delta T(\vec{k},\eta) = \Theta(\vec{k},\eta) P(\vec{k},\eta)+G(\vec{k},\eta)$, where $P(\vec{k},\eta)$
is the mode average of $P_Q(\vec{k},\eta)$, and $G(\vec{K},\eta)$ vanishes if $P$ is independent of mode $Q$.
This paper simplifies by choosing $P$ independent of $Q$.

\subsection{1D heat transport in dc limit}

We now focus on the dc  heat transport along the $x$ direction of graphene.
Therefore $\eta=0$, and the wavevector $\vec{k}$ and velocity $\vec{v}_Q$
have only one relevant component, $k_x\equiv k$, and  $v_{Qx} \equiv v_Q$. 
The thermal distributor simplifies to
\begin{equation}
    \Theta_{\rm LPBE}(k) = \frac{1}{C}\frac{\langle X| \hat{W} ^{-1}|X\rangle}{\langle X| \hat{W} ^{-1}|ikvX\rangle}
\label{Equ:ThetaLPBE}
\end{equation}
We label it LPBE because it is obtained from the linear PBE Eq.~(\ref{Equ:PBEWX}), and we want to distinguish it from
the RTA version of the PBE.  The local temperature $\Delta T(\vec{r})$ 
appearing in the LPBE, Eq. \ref{Equ:EC}, is defined by the statement that
the local equilibrium distribution $n_Q(T(\vec{r}))$ carries all the heat, and the deviation $N_Q-n_Q(T(\vec{r}))$
carries no heat; $\Delta U(\vec{r})$ in Eq. \ref{Equ:EC} is zero.  How is $\Delta T(\vec{r})$ defined in RTA?
It is a peculiar fact of the RTA that an alternative definition of $\Delta T(\vec{r})$ is preferable, namely 
\begin{eqnarray}
\left( \frac{\partial U}{\partial t}\right)_{\rm scat}^{\rm RTA} &=& 
\sum_Q \hbar\omega_Q \left(\frac{\partial N_Q}{\partial t} \right)_{\rm scat}^{\rm RTA}
\nonumber \\
=0=&-&\sum_Q \hbar\omega_Q \frac{N_Q-n_Q(T(\vec{r},t))}{\tau_Q} \nonumber \\
\label{eq:RTAEC}
\end{eqnarray}
This option is known to work better than the alternative of setting $\Delta U$ to 0 \cite{allen2018temperature}.  
The RTA result for the thermal distributor is then
\begin{equation}
    \Theta_{\rm RTA}(k) = \frac{1}{C} \frac{\sum_Q\frac{C_Q\Gamma_Q^2}
    {\Gamma_Q^2+(kv_{Q})^2}}
    {\sum_Q\frac{C_Q\Gamma_Q(kv_{xQ})^2}
    {\Gamma_Q^2+(kv_{Q})^2}}
\label{Equ:ThetaRTA}
\end{equation}
where $\Gamma_Q=1/\tau_Q$.  


\section{\label{sec:level1}Results and Discussion}

\begin{figure*}
  \centering
 \includegraphics[width=1\textwidth]{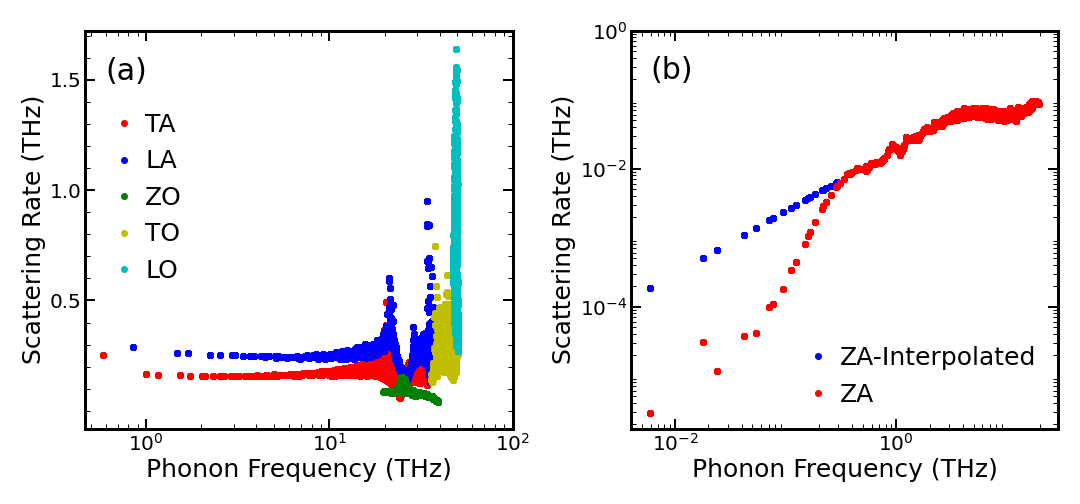}
 \caption{Anharmonic three-phonon scattering rates for graphene at room temperature are shown in (a) on a linear scale; 
 (b) shows the scattering rates of ZA modes on a log scale. The blue circles show linear interpolation of the ZA scattering rate for frequencies below 0.3 THz, see text.}
\label{Fig:SR}
\end{figure*}

Quasiparticle heat transport in solids can be roughly categorized into three regimes: 
ballistic, quasi-ballistic, and diffusive. In the diffusive regime, where Fourier's law applies, the channel length of the heat conductors 
is much longer than the MFPs; phonons experience multiple scattering events, so that a local equilibrium is 
established, with a temperature gradient that is constant everywhere except in the small region close to heat sources and sinks. 
Thermal transport is ballistic when the channel length is comparable to or less than phonon MFPs. 
In this case, thermal energy is mainly dissipated near the heat sources.  The temperature gradient 
becomes thermally inhomogeneous, and heat current has a non-local relation to temperature.
When the channel length is similar to the averaged phonon MFP, some phonons travel ballistically and others diffusely;
heat transfer is ``quasi-ballistic''.  
The wide range of phonon MFPs in graphene ~\cite{li2019crossover} makes it hard to differentiate transport regimes. 
Moreover, the heat transport regime in a given device is temperature-dependent 
since the phonon MFPs depend on temperature.

The thermal distributor $\Theta(\vec{r})$, Eq. \ref{eq:deftheta}, simplifies the description of heat transport in different regimes. 
The spatial variation of temperature is given by
\begin{equation}
T(\vec{r})=\sum_{\vec{k}}T(\vec{k})e^{i\vec{k}\cdot\vec{r}}=\sum_{\vec{k}} \Theta(\vec{k})P(\vec{k})e^{i\vec{k}\cdot\vec{r}}.
\label{eq:TfromTheta}
\end{equation}
The Fourier transform $\Theta(\vec{k})$ is related to the non-local generalization of the bulk thermal conductivity 
$\kappa_{\rm bulk} = {\rm lim}_{\vec{k}\rightarrow 0} \kappa(\vec{k})$ by Eq. \ref{Equ:ThetaKappa}.
First, we calculate LPBE and RTA versions of $\Theta(\vec{k})$ of graphene from Eq.~(\ref{Equ:ThetaLPBE}) 
and Eq.~(\ref{Equ:ThetaRTA}), using the modified Tersoff potential, with the parameters given in ref. \onlinecite{lindsay2010optimized}, 
to model the crystal potential. The scattering rates are shown in Fig.~\ref{Fig:SR} at $T=300$ K.

Note that there are numerical challenges in calculating the scattering rates of phonons using the Gaussian broadening~\cite{Gu2019}. 
Following Ref.~\cite{lindsay2014phonon,Bonini2012}, we apply linear interpolation of the ZA scattering rates with phonon energy, as shown by the blue circles for phonon energies below 0.3 THz in Fig.~{\ref{Fig:SR}}b. This linear scaling is explained by noting that the bending energy is given by $E_{b}=\int d^2r \kappa_b/(2R^2)$, where $\kappa_b=2.1$ eV~\cite{Valencemodel2009} is the bending stiffness, and $R$ is the radius of curvature given by $1/R=d^2z/dx^2$. Applying the Bloch theorem for a discrete atomic model with a lattice constant $a$, one can show that bending energy for mode $q$ is given by $ E_{b}=8\kappa_b A_c z_q^2\sin^4(qa/2)/a^4\approx A_c\kappa_b z_q^2 q^4/2$, where $A_c$ is the area per atom. By comparing it to the harmonic oscillator potential energy $m\omega_q^2z_q^2/2$, one can obtain the expected result for flexural phonon frequency: $\omega_q=q^2\sqrt{\kappa_bA_c/m}$. The third order anharmonicity can be introduced by coupling the ZA mode to an LA mode by modifying the bending stiffness $\kappa_b=\kappa_{b0}-\alpha_b(x_{i+1}-x_{i-1})$. After applying the Bloch theorem, the third-order anharmonic potential for mode $q$ in the small $q$-limit becomes $H_3=\sum_{q_1}i\alpha_bA_cz_qz_{q_1}x_{-q-q_1}q^2q_1^2(q+q_1)a$. Using second-quantized amplitudes for the phonon displacements $z_q$ and $x_q$, one can show that $V_{qq_1}\sim q^2q_1^2(q+q_1)(\omega_{q}^{ZA}\omega_{q_1}^{ZA}\omega_{q+q_1}^{LA})^{-1/2}$. The $q$-ZA phonon scattering with $q_1$-ZA phonon into an $(q+q_1)$-LA phonon has a rate of $1/\tau^{ZA}_{q}\sim \sum_{q_1} \vert V_{q,q_1}\vert^2n^{0, ZA}_{q_1}n^{0, LA}_{q+q_1}\delta(\omega^{ZA}_q+\omega^{ZA}_{q_1}-\omega^{LA}_{q+q_1})$. The $\delta$-function ensures that $q_1\sim v_s\sqrt{m/(\kappa_bA_c)}$, where $v_s$ is the sound velocity. Therefore, the interpolation function $1/\tau^{ZA}_{q}\sim q^2\sim \omega^{ZA}_{q}$ used in Fig.~\ref{Fig:SR}b can be justified.

\begin{figure}
 \includegraphics[width=0.5\textwidth]{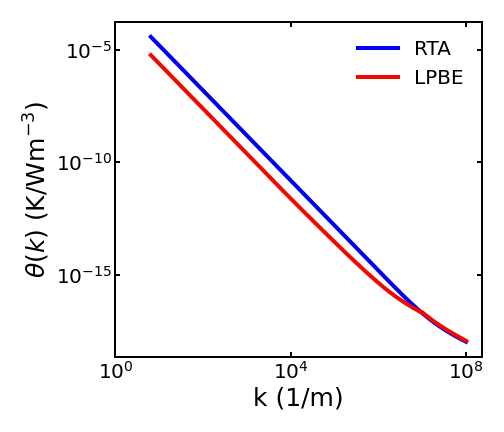}
 \caption{Fourier transform of the thermal distributor as a function of momentum $k$ in graphene using RTA and LPBE approaches. The small $k$ limit corresponds to large distances from the input of the heat source such that thermal transport is diffusive and $\Theta(k)$ diverges.}
\label{Fig:Thetak}
\end{figure}

\begin{figure*}
  \centering
   \includegraphics[width=1\textwidth]{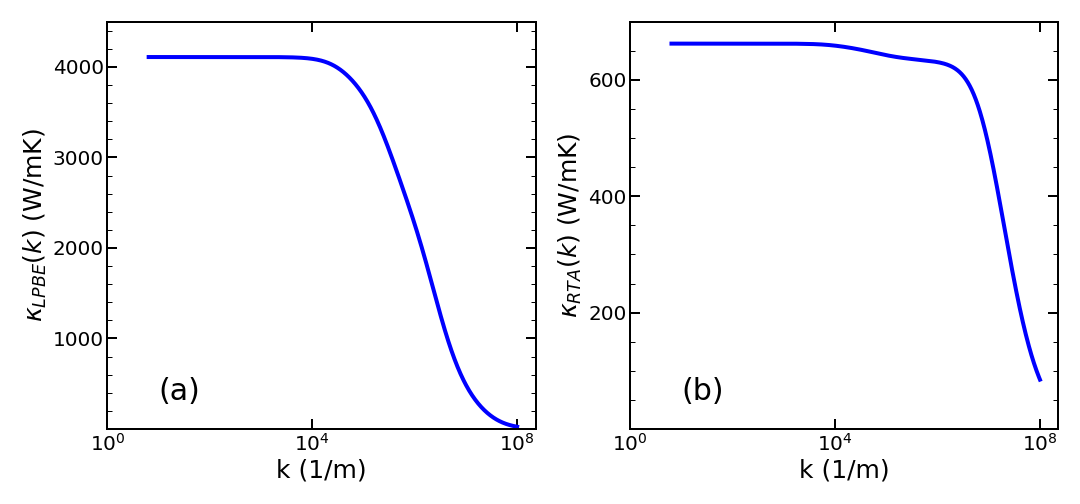}
 \caption{Thermal conductivity using LPBE (a) and RTA (b) approaches.}
\label{Fig:Kappa(k)}
\end{figure*}

Fig.~\ref{Fig:Thetak}  shows the calculated $\Theta(k)$ for graphene using LPBE and RTA approximations. 
The width of the sample is much broader than the MFP; this allows a one-dimensional treatment.  
The spatial variation $\vec{r}=(x,0)$  is only along $\hat{x}$, parallel to the channel, so the spatial Fourier
variable is $\vec{k}=(k,0)$.  The spatial resolution of $T(x)$ 
at small distances $x$ requires values of $\Theta(k)$ at correspondingly large $k \sim 2\pi/x$
(see Eq. \ref{eq:TfromTheta}).  The range of $k$ in Fig.~\ref{Fig:Thetak} corresponds to distances 
from a few nanometers to a meter-long channel length.  A lower limit of the length scale is imposed by 
the validity of the quasiparticle picture of phonons used in the PBE formalism ~\cite{allen2018analysis}. 
Note that the thermal distributor diverges for both LPBE and RTA solutions when $k\to0$. 
According to Eq.~(\ref{Equ:ThetaKappa}), for a finite thermal conductivity in a diffusive regime, 
$\Theta(k)$ must diverge in $k\to0$ limit as  $\Theta(k)\sim\frac{1}{k^2}$.   Curve-fitting 
shows that the calculated $\Theta(k)$ for both versions agrees with  $\frac{1}{k^2}$ very accurately in the 
small $k$ limit, namely:
\begin{equation}
\begin{split}
   &  \Theta_{LBPE}(k) = \frac{2.4\times10^{-4} \ {\rm Km/W }}{k^2}  \\
   &  \Theta_{RTA}(k) = \frac{1.51\times10^{-3} \ {\rm Km/W }}{k^2} 
   \label{Edu:theta_fit}
\end{split}
\end{equation}
Using Eq.~(\ref{Equ:ThetaKappa}), this corresponds to bulk thermal conductivities  4145 W/mK and 662 W/mK 
for LPBE and RTA, respectively. The large discrepancy between LPBE and RTA values of
thermal conductivities of graphene has been noticed previously \cite{lindsay2014phonon}. 


The thermal conductivities evaluated according to Eq.~(\ref{Equ:Appx5}) for LPBE and Eqs.~\ref{Equ:ThetaKappa}, \ref{Equ:ThetaRTA} for RTA are shown in Fig.~\ref{Fig:Kappa(k)}. 
The sharp fall-off of the thermal conductivity in  Fig.~\ref{Fig:Kappa(k)} 
indicates the ballistic-to-diffusive crossover; it happens at larger $k$, 
and, therefore, smaller characteristic lengths in the RTA treatment than in the correct LPBE treatment. 

\begin{figure*}
  \centering
 \includegraphics[width=1\textwidth]{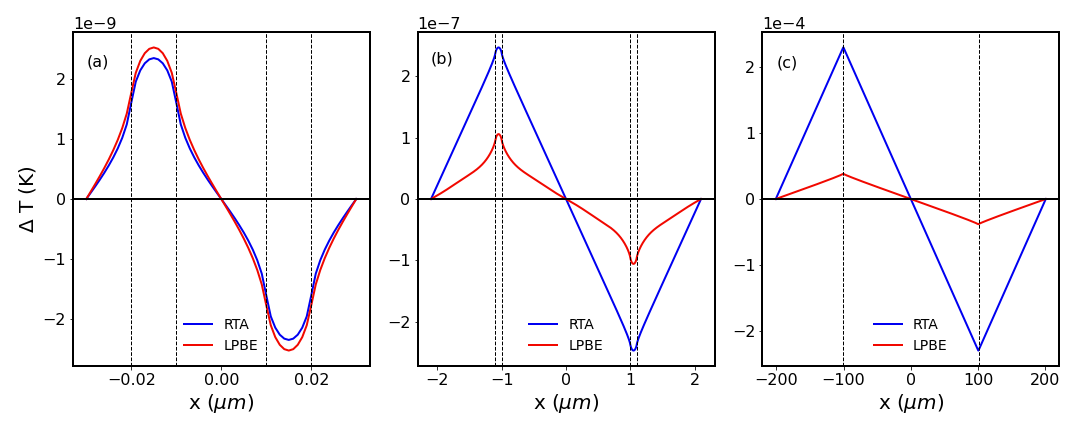}
 \caption{Temperature variation $\Delta T(x)=T(x)-T_0$ at $T_0$=300K
 for three different values of the channel length: (a) $L_{\rm ch}=20$ nm, (b) $L_{\rm ch}=
 2 \ \mu$m, and (c) $L_{\rm ch}= 200 \ \mu$m. The source/sink lengths are $L_{\rm s}=10$ nm, 
 100 nm, 1 $\mu$m in (a), (b), and (c), correspondingly.
 Input power $P(x)=P_0 = (1 {\rm W/m^2})/h= 2.94\times 10^9 {\rm W/m^3}$ is applied in all calculations.  The effective thermal conductivities are given in table~\ref{table:I}.}
\label{Fig:T-Profile}
\end{figure*}

Now we discuss thermal conduction in the geometry of Fig. \ref{fig:graphenemodel} using the PBE to evaluate the 
thermal distributor.  
We can calculate the temperature profile for any pattern of heat input and removal from this response function, provided the graphene sample has one-dimensional periodicity. Energy conservation requires $dJ/dx=-P(x)$.   Steady-state
($\eta=0$) requires equal external heat addition and removal. 
$L_{\rm ch}$ is the length of the channel between the two heat reservoirs, source and sink; each of length $L_{\rm s}$. 
For simplicity, our heat input has odd symmetry ($P(-x)=-P(x)$) around $x=0$,
so $J$ is even in $x$.  The period of the supercell is $L_d=2(L_{\rm ch}+L_{\rm s})$, which determines the shortest non-zero wavevector i.e. $k_{min} = 2\pi/L_d$. A large $L_d$ value allows a fine Fourier mesh to describe nanoscale physics, such as the ballistic-to-diffusive crossover.
Our reservoirs are ideal thermal baths, with zero interfacial thermal resistance 
between the channel and reservoirs. Note that experimental thermal conductivity measurements are often 
done using periodic structures such as periodic metallic gratings or transient thermal gratings ~\cite{johnson2013direct, 
zeng2015measuring}.  Our periodic geometry of Fig. \ref{fig:graphenemodel} works for both periodic structures and
single-channel devices.  In the latter case, it is necessary to make $L_{\rm s}$ larger than phonon 
MFPs.

Figure \ref{fig:graphenemodel} shows how heat insertion and removal $P(x)$ is distributed uniformly 
(with magnitude $P_0$) 
over lengths $L_{\rm s}$ on either side of the sample (or channel) of length $L_{\rm ch}$.  
Energy conservation $dJ/dx=P$
then gives heat current density $J=P_0 L_{\rm s}/2$ in the channels.
Figure \ref{Fig:T-Profile} shows the resulting  $\Delta T$ profiles in three devices 
with channel lengths spanning from ballistic to diffusive regimes.
In the ballistic device, Fig.~\ref{Fig:T-Profile}a, both 
$\Theta_{RTA}$ and $\Theta_{LBTE}$ predict similar values for temperature profiles in the channel and in the source/sink reservoirs.  This behavior can be understood from the two thermal 
distributors being similar in magnitude in the large $k$-limit, as shown in  Fig.~\ref{Fig:Thetak}.  The temperature profile of 
Fig.~\ref{Fig:T-Profile}a clearly shows the source regions are hotter than the channel, which is a characteristic 
of non-diffusive thermal transport~\cite{chiloyan2020thermal, hua2018heat, li2019influence}. The long MFP phonons (ZA phonons) dissipate the thermal energy in the source regions while flying in the channel ballistically. As a result, the temperature is higher in the heat region. This observation manifests the nonlocal effect.

For the larger channel lengths in Fig.~\ref{Fig:T-Profile}b and \ref{Fig:T-Profile}c,
RTA predictions for the temperature are significantly higher than LPBE predictions in agreement with 
Fig. \ref{Fig:Thetak} which shows
that at smaller $k$ (corresponding to larger distances)
$\Theta_{RTA}$ is greater than $\Theta_{LPBE}(k)$. 
In the quasi-ballistic regime of Fig.~\ref{Fig:T-Profile}b, there is significant non-linearity near the heat source/sink. 
Although in this regime both long and short MFP phonons contribute to thermal transport, the share of thermal energy carried by short MFP phonons is almost negligible.


Non-local heat transport is often described by an ``effective thermal conductivity'' $\kappa_{\rm eff}$.  The definition
varies depending on the experiment.  For example,
experimental studies such as 
time-domain thermoreflectance \cite{Cahill2004} measure the transient temperature response
to a heat pulse.   This can be used to extract $\kappa_{\rm eff}$ \cite{johnson2013direct}. 
Many theoretical studies also have applied a similar procedure. 

\begin{table}
\begin{center}
\begin{tabular}{|c|c|c|c|c| } 
 \hline
LPBE & definition & Fig. \ref{Fig:T-Profile}a   & Fig. \ref{Fig:T-Profile}b    & Fig. \ref{Fig:T-Profile}c    \\ 
 \hline
 $\kappa_{\rm eff, mp}$ & $\frac{J_0}{(dT/dx)_{x=0}}$ & 118 & 2463 & 4035 \\ 
 $\kappa_{\rm eff, min}$ & $\frac{J_0}{ \Delta T_{\rm max}/(L_{\rm ch}+L_{\rm s})}$   & 87 & 1460 & 3854 \\ 
 $\kappa_{\rm eff, ch}$ & $\frac{J_0}{ \Delta T_{\rm ch}/L_{\rm ch}}$  & 124 & 1565 & 3860 \\  
 \hline
 RTA &definition & Fig. \ref{Fig:T-Profile}a   & Fig. \ref{Fig:T-Profile}b    & Fig. \ref{Fig:T-Profile}c    \\ 
 \hline 
 $\kappa_{\rm eff, mp}$ & $\frac{J_0}{ (dT/dx)_{x=0}}$ & 139 & 633 & 658 \\ 
 $\kappa_{\rm eff, min}$ & $\frac{J_0}{ \Delta T_{\rm max}/(L_{\rm ch}+L_{\rm s})}$   & 85 & 617 & 656 \\ 
 $\kappa_{\rm eff, ch}$ & $\frac{J_0}{ \Delta T_{\rm ch}/L_{\rm ch}}$  & 116 & 616 & 654 \\  
 \hline
 \end{tabular}
\caption{\label{table:I} Effective thermal conductivities computed from three definitions for
the cases shown in Fig. \ref{Fig:T-Profile}.  The subscript `mp' means 
current density divided by mid-point temperature gradient; `min' means current divided by maximum $\Delta T$,
per half supercell length; and `ch' means current divided by the temperature difference at the channel edges, per channel length.
The current density $J_0$ is the constant value in the channel.}
\end{center}
\end{table}
%

\begin{figure*}
  \centering
  \includegraphics[width=1\textwidth]{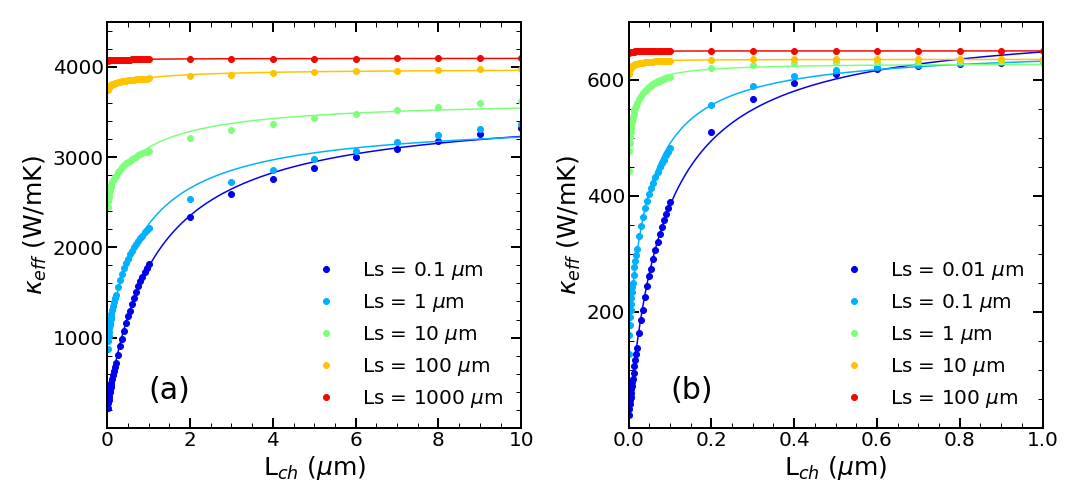}
  \hfill
 \caption{Effective mid-point thermal conductivity at 300K versus channel length calculated from the PBE
 and the geometry of Fig. \ref{fig:graphenemodel}, for different source lengths: 
 (a) LPBE and (b) RTA solutions. For $L_{\rm s}\leq 100 \mu m$ ($L_{\rm s}\leq 10 \mu m$) 
 the transition from ballistic or quasi-ballistic to diffusive can be observed in LPBE (RTA) solutions. 
 For $L_{\rm s}>100 \mu m$ ( $L_{\rm s}>10 \mu m$) the thermal transport is always in the diffusive regime 
 for LPBE (RTA) solution regardless of the source length. The solid curves are best fits to
 $\kappa_{\rm eff,a}$ in Eq.~(\ref{Edu:Kappa_fit_a}).}
\label{Fig:Keff(Lch)}
\end{figure*}
%

\begin{figure*}
  \centering
  \includegraphics[width=1\textwidth]{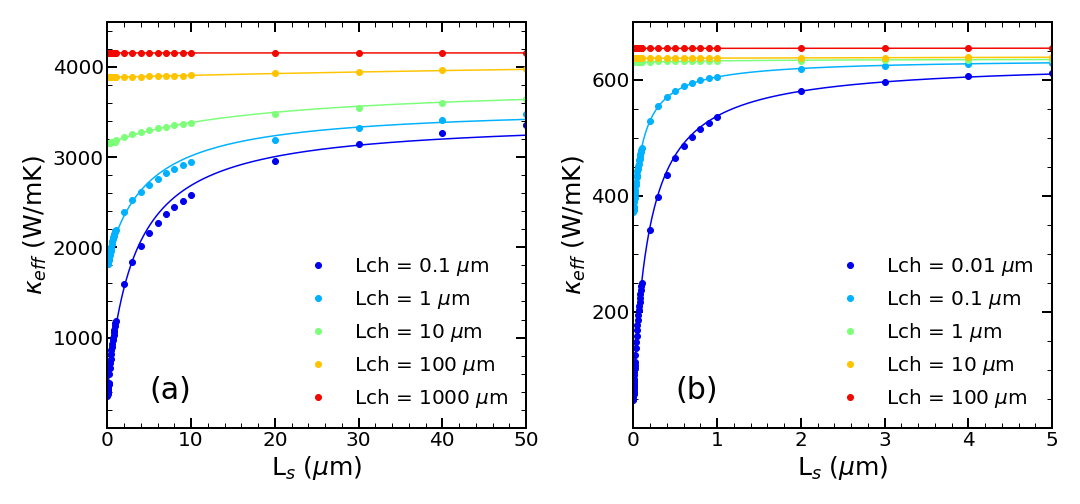}
 \caption{Effective mid-point thermal conductivity at 300K versus source length calculated for different channel lengths: 
 (a) LPBE and (b) RTA solutions. For $L_{\rm ch}\leq 100 \mu m$ ($L_{\rm ch}\leq 10 \mu m$), 
 the transition from ballistic or quasi-ballistic to diffusive can be observed in LPBE (RTA) solutions. 
 For $L_{\rm ch}>100 \mu m$ ( $L_{\rm ch}>10 \mu m$) the thermal transport is always in the diffusive regime 
 for LPBE (RTA) solution regardless of the source length. The solid curves are best fits to 
 $\kappa_{\rm eff,b}$ in Eq.~(\ref{Edu:Kappa_fit_b}).}
\label{Fig:Keff(Ls)}
\end{figure*}

For our steady-state thermal transport computations, there are three sensible
definitions of $\kappa_{\rm eff}$, shown in table~\ref{table:I}.  The first of these defines $\kappa_{\rm eff}$ 
by applying Fourier's law at the middle of the channel where the curvature of 
$\Delta T(x)$ is zero. This resembles
Fourier's law in thermally homogeneous structures. We note that in the fully diffusive regime, when both $L_{\rm s}$ and $L_{\rm ch}$ are larger than the MFPs, the $\kappa_{\rm eff, mp}$ and $\kappa_{\rm eff, ch}$ values coincide, and the temperature drop in the channel is $\Delta T_{\rm ch}=J_0L_{\rm ch}/\kappa_{\rm eff, mp}$. In this limit, the temperature drop under the contact is equal to $J_0L_{\rm s}/(4\kappa_{\rm eff, mp})$, because current $J(x)$ varies linearly with distance from the middle of the contact. Therefore, the maximum temperature variation across the unit cell equals $\Delta T_{\rm max}=\Delta T_{\rm ch}+J_0L_{\rm s}/(2\kappa_{\rm eff, mp})$ and effective thermal conductivity $\kappa_{\rm eff, min}=J_0(L_{\rm ch}+L_{\rm s})/\Delta T_{\rm max}=\kappa_{\rm eff, mp} (L_{\rm ch}+L_{\rm s})/ (L_{\rm ch}+L_{\rm s}/2)$. As can be seen from table~\ref{table:I}, the device geometry in Fig.~\ref{Fig:T-Profile}c approaches the diffusive limit. However, one should note the fully diffusive limit relationship between  $\kappa_{\rm eff, min}$ and $\kappa_{\rm eff, mp}$ is not realized in the devices we considered.

The $\kappa_{\rm eff, mp}$ definition is plotted in Fig.~\ref{Fig:Keff(Lch)} and \ref{Fig:Keff(Ls)}, for both LPBE and RTA 
versions of the thermal distributor, as a function of the source length and channel length, respectively. 
When the chosen length $L_{\rm ch}$ in Fig. \ref{Fig:Keff(Lch)} (or $L_{\rm s}$ in Fig. \ref{Fig:Keff(Ls)}) 
has a value less than the diffusive length scale ($\sim 1 -3\: \mu$m for LPBE and $\sim 100-200\:$ nm for RTA),  
$\kappa_{\rm eff, mp}$ decreases with decreasing periodicity due to suppression of the ballistic phonon 
contribution to the heat transport.  At large values of $L_{\rm ch}$, thermal conductivity saturates and approaches the diffusive limit when both $L_{\rm ch}$ and $L_{\rm s}$ are large. 

Similarly, Fig. \ref{Fig:Keff(Ls)} shows that $\kappa_{\rm eff, mp}$ saturates with increasing $L_{\rm s}$ and reaches the diffusive limit value at large $L_{\rm ch}$. For small $L_{\rm ch}$ values,  the $\kappa_{\rm eff, mp}$ dependence on $L_{\rm s}$ shows ballistic to diffusive crossover as in Fig.~\ref{Fig:Keff(Lch)}. 
Those dependencies happen because of superposing interaction of ballistic phonons in the ballistic channels and recovery of diffusive-like thermal transport. This phenomenon shows the deterministic role of the geometry of the heat source and has already been observed experimentally ~\cite{zeng2015measuring,hoogeboom2015new}.

To quantify our results in Fig.~\ref{Fig:Keff(Lch)} and \ref{Fig:Keff(Ls)}, we fit our $\kappa_{\rm eff, mp}$ to the phenomenological 
ballistic-to-diffusive crossover equations used to describe electrical transport~\cite{lundstrom_2000}: 
\begin{eqnarray}
    \kappa_{\rm eff,a} &=& \kappa_0+(\kappa_{\rm dif}-\kappa_0)\frac{L_{\rm ch}}{L_{\rm ch}+\lambda}  \
   \label{Edu:Kappa_fit_a} \\
   \kappa_{\rm eff,b} &=& \kappa_0+(\kappa_{\rm dif}-\kappa_0)\frac{L_{\rm s}}{L_{\rm s}+\lambda}
   \label{Edu:Kappa_fit_b}
\end{eqnarray}
where $\kappa_{\rm dif}$ is thermal conductivity in the diffusive regime. From the best fits, we find a characteristic length scale $\lambda\sim 1$ $\mu$m ($\sim 90$ nm) for the LPBE (RTA) solution using fits of $\kappa_{\rm eff,a}$ versus $L_{\rm ch}$, and $\lambda\sim 3$ $\mu$m ($\sim 200$ nm) 
for LPBE (RTA) solution using fits of $\kappa_{\rm eff,b}$ versus $L_{\rm s}$. The value of $\kappa_0$ 
approaches zero only when both $L_{\rm s}$ and $L_{\rm ch}$ are small, as discussed above.  
Those estimates for the characteristic crossover length scales are consistent with the $k$-dependences 
of thermal conductivity in Fig.~\ref{Fig:Kappa(k)}. Using a characteristic value of $k_0$, 
when thermal conductivity drops by a factor of two in Fig.~\ref{Fig:Kappa(k)}, we find 
$2\pi/k_0\sim 5$ $\mu$m for LPBE and  $2\pi/k_0\sim 300$ nm for RTA. 
Finally, we can estimate an average phonon MFP using the standard expression 
for thermal conductivity in 2D: $\kappa=Cv\lambda_{\rm ph}/2$, where the heat capacity is  
$C\times h=4.5\times10^{-4} {\rm J/Km}^2$ (calculated at $T=300$ K)
and $v$ is an averaged phonon velocity. Using the velocity of the ZA parabolic band 
at room temperature $v\approx 10$ km/s, 
we obtain $\lambda_{\rm ph}=625$ nm for LPBE and $\lambda_{\rm ph}=100$ nm for RTA, consistent with 
the above estimates for the diffusive to ballistic crossover length scales.

\section{\label{sec:level1}Conclusions}

Steady microscale addition and removal of heat energy, together with a scattering of heat carriers, leads to carrier distributions $N$ centered around  local equilibrium $n$ with a local temperature  $T(x)$. 
The thermal distributor $\Theta(k)$ contains all relevant information about thermal transport in all the regimes.
By computing the thermal distributor $\Theta(k)$ of graphene from the PBE with correct inversion using the full
scattering operator, we investigated thermal transport 
at submicron length scales. At long length scales, heat transport occurs with constant temperature gradients. 
Non-local effects (where $\nabla T$ varies with $x$) are seen at length scales of the order of or less than 
mean free paths of phonons.   Details of the geometrical structure of the device 
and heat sources and sinks cause the inhomogeneous temperature profile $T(x)$. 
The long-wavelength phonons are forcefully scattered in the source and sink regions while flying the channel without appreciable scatterings with other phonons.
This causes local $\Delta T(x)$ to be higher near the source/sink.  This is often ascribed to the suppression of heat transport by phonons with long mean free paths. The RTA contains these effects but,
especially in graphene, overestimates the temperature inhomogeneity needed to drive a heat current.

\section*{Acknowledgments}

We acknowledge support from the Vice President for Research and Economic Development (VPRED), SUNY Research Seed Grant Program, and the Center for Computational Research at the University at Buffalo~\cite{UBCCR}.


\bibliography{reference.bib}

\begin{thebibliography}{48}%
\makeatletter
\providecommand \@ifxundefined [1]{%
 \@ifx{#1\undefined}
}%
\providecommand \@ifnum [1]{%
 \ifnum #1\expandafter \@firstoftwo
 \else \expandafter \@secondoftwo
 \fi
}%
\providecommand \@ifx [1]{%
 \ifx #1\expandafter \@firstoftwo
 \else \expandafter \@secondoftwo
 \fi
}%
\providecommand \natexlab [1]{#1}%
\providecommand \enquote  [1]{``#1''}%
\providecommand \bibnamefont  [1]{#1}%
\providecommand \bibfnamefont [1]{#1}%
\providecommand \citenamefont [1]{#1}%
\providecommand \href@noop [0]{\@secondoftwo}%
\providecommand \href [0]{\begingroup \@sanitize@url \@href}%
\providecommand \@href[1]{\@@startlink{#1}\@@href}%
\providecommand \@@href[1]{\endgroup#1\@@endlink}%
\providecommand \@sanitize@url [0]{\catcode `\\12\catcode `\$12\catcode
  `\&12\catcode `\#12\catcode `\^12\catcode `\_12\catcode `\%12\relax}%
\providecommand \@@startlink[1]{}%
\providecommand \@@endlink[0]{}%
\providecommand \url  [0]{\begingroup\@sanitize@url \@url }%
\providecommand \@url [1]{\endgroup\@href {#1}{\urlprefix }}%
\providecommand \urlprefix  [0]{URL }%
\providecommand \Eprint [0]{\href }%
\providecommand \doibase [0]{https://doi.org/}%
\providecommand \selectlanguage [0]{\@gobble}%
\providecommand \bibinfo  [0]{\@secondoftwo}%
\providecommand \bibfield  [0]{\@secondoftwo}%
\providecommand \translation [1]{[#1]}%
\providecommand \BibitemOpen [0]{}%
\providecommand \bibitemStop [0]{}%
\providecommand \bibitemNoStop [0]{.\EOS\space}%
\providecommand \EOS [0]{\spacefactor3000\relax}%
\providecommand \BibitemShut  [1]{\csname bibitem#1\endcsname}%
\let\auto@bib@innerbib\@empty
\bibitem [{\citenamefont {Peierls}(1929)}]{Peierls29}%
  \BibitemOpen
  \bibfield  {author} {\bibinfo {author} {\bibfnamefont {R.}~\bibnamefont
  {Peierls}},\ }\bibfield  {title} {\bibinfo {title} {{Zur kinetischen Theorie
  der Wärmeleitung in Kristallen}},\ }\href@noop {} {\bibfield  {journal}
  {\bibinfo  {journal} {Annalen der Physik}\ }\textbf {\bibinfo {volume}
  {395}},\ \bibinfo {pages} {1055} (\bibinfo {year} {1929})}\BibitemShut
  {NoStop}%
\bibitem [{\citenamefont {Simons}(1960)}]{simons1960boltzmann}%
  \BibitemOpen
  \bibfield  {author} {\bibinfo {author} {\bibfnamefont {S.}~\bibnamefont
  {Simons}},\ }\bibfield  {title} {\bibinfo {title} {{The Boltzmann equation
  for a bounded medium I. General theory}},\ }\href@noop {} {\bibfield
  {journal} {\bibinfo  {journal} {Phil. Trans. Roy. Soc. London. Series A,
  Math. Phys. Sci.}\ }\textbf {\bibinfo {volume} {253}},\ \bibinfo {pages}
  {137} (\bibinfo {year} {1960})}\BibitemShut {NoStop}%
\bibitem [{\citenamefont {Levinson}(1980)}]{levinson1980nonlocal}%
  \BibitemOpen
  \bibfield  {author} {\bibinfo {author} {\bibfnamefont {Y.}~\bibnamefont
  {Levinson}},\ }\bibfield  {title} {\bibinfo {title} {Nonlocal phonon heat
  transfer},\ }\href@noop {} {\bibfield  {journal} {\bibinfo  {journal} {Sol.
  State Commun.}\ }\textbf {\bibinfo {volume} {36}},\ \bibinfo {pages} {73}
  (\bibinfo {year} {1980})}\BibitemShut {NoStop}%
\bibitem [{\citenamefont {Mahan}\ and\ \citenamefont
  {Claro}(1988)}]{mahan1988nonlocal}%
  \BibitemOpen
  \bibfield  {author} {\bibinfo {author} {\bibfnamefont {G.}~\bibnamefont
  {Mahan}}\ and\ \bibinfo {author} {\bibfnamefont {F.}~\bibnamefont {Claro}},\
  }\bibfield  {title} {\bibinfo {title} {Nonlocal theory of thermal
  conductivity},\ }\href@noop {} {\bibfield  {journal} {\bibinfo  {journal}
  {Phys. Rev. B}\ }\textbf {\bibinfo {volume} {38}},\ \bibinfo {pages} {1963}
  (\bibinfo {year} {1988})}\BibitemShut {NoStop}%
\bibitem [{\citenamefont {Majumdar}(1993)}]{majumdar1993microscale}%
  \BibitemOpen
  \bibfield  {author} {\bibinfo {author} {\bibfnamefont {A.}~\bibnamefont
  {Majumdar}},\ }\bibfield  {title} {\bibinfo {title} {Microscale heat
  conduction in dielectric thin films},\ }\href@noop {} {\bibfield  {journal}
  {\bibinfo  {journal} {ASME J. Heat Transfer}\ }\textbf {\bibinfo {volume}
  {115}},\ \bibinfo {pages} {7} (\bibinfo {year} {1993})}\BibitemShut {NoStop}%
\bibitem [{\citenamefont {Chen}(2005)}]{chen2005nanoscale}%
  \BibitemOpen
  \bibfield  {author} {\bibinfo {author} {\bibfnamefont {G.}~\bibnamefont
  {Chen}},\ }\href@noop {} {\emph {\bibinfo {title} {Nanoscale energy transport
  and conversion: a parallel treatment of electrons, molecules, phonons, and
  photons}}}\ (\bibinfo  {publisher} {{Oxford University Press}},\ \bibinfo
  {year} {2005})\BibitemShut {NoStop}%
\bibitem [{\citenamefont {Ordonez-Miranda}\ \emph {et~al.}(2011)\citenamefont
  {Ordonez-Miranda}, \citenamefont {Yang},\ and\ \citenamefont
  {Alvarado-Gil}}]{ordonez2011constitutive}%
  \BibitemOpen
  \bibfield  {author} {\bibinfo {author} {\bibfnamefont {J.}~\bibnamefont
  {Ordonez-Miranda}}, \bibinfo {author} {\bibfnamefont {R.}~\bibnamefont
  {Yang}},\ and\ \bibinfo {author} {\bibfnamefont {J.}~\bibnamefont
  {Alvarado-Gil}},\ }\bibfield  {title} {\bibinfo {title} {A constitutive
  equation for nano-to-macro-scale heat conduction based on the {Boltzmann}
  transport equation},\ }\href@noop {} {\bibfield  {journal} {\bibinfo
  {journal} {J. Appl. Phys.}\ }\textbf {\bibinfo {volume} {109}},\ \bibinfo
  {pages} {084319} (\bibinfo {year} {2011})}\BibitemShut {NoStop}%
\bibitem [{\citenamefont {Allen}\ and\ \citenamefont
  {Perebeinos}(2018)}]{allen2018temperature}%
  \BibitemOpen
  \bibfield  {author} {\bibinfo {author} {\bibfnamefont {P.~B.}\ \bibnamefont
  {Allen}}\ and\ \bibinfo {author} {\bibfnamefont {V.}~\bibnamefont
  {Perebeinos}},\ }\bibfield  {title} {\bibinfo {title} {{Temperature in a
  Peierls-Boltzmann treatment of nonlocal phonon heat transport}},\ }\href@noop
  {} {\bibfield  {journal} {\bibinfo  {journal} {Phys. Rev. B}\ }\textbf
  {\bibinfo {volume} {98}},\ \bibinfo {pages} {085427} (\bibinfo {year}
  {2018})}\BibitemShut {NoStop}%
\bibitem [{\citenamefont {Hua}\ \emph {et~al.}(2019)\citenamefont {Hua},
  \citenamefont {Lindsay}, \citenamefont {Chen},\ and\ \citenamefont
  {Minnich}}]{hua2019generalized}%
  \BibitemOpen
  \bibfield  {author} {\bibinfo {author} {\bibfnamefont {C.}~\bibnamefont
  {Hua}}, \bibinfo {author} {\bibfnamefont {L.}~\bibnamefont {Lindsay}},
  \bibinfo {author} {\bibfnamefont {X.}~\bibnamefont {Chen}},\ and\ \bibinfo
  {author} {\bibfnamefont {A.~J.}\ \bibnamefont {Minnich}},\ }\bibfield
  {title} {\bibinfo {title} {{Generalized Fourier's law for nondiffusive
  thermal transport: Theory and experiment}},\ }\href@noop {} {\bibfield
  {journal} {\bibinfo  {journal} {Phys. Rev. B}\ }\textbf {\bibinfo {volume}
  {100}},\ \bibinfo {pages} {085203} (\bibinfo {year} {2019})}\BibitemShut
  {NoStop}%
\bibitem [{\citenamefont {Hua}\ and\ \citenamefont
  {Lindsay}(2020)}]{hua2020space}%
  \BibitemOpen
  \bibfield  {author} {\bibinfo {author} {\bibfnamefont {C.}~\bibnamefont
  {Hua}}\ and\ \bibinfo {author} {\bibfnamefont {L.}~\bibnamefont {Lindsay}},\
  }\bibfield  {title} {\bibinfo {title} {Space-time dependent thermal
  conductivity in nonlocal thermal transport},\ }\href@noop {} {\bibfield
  {journal} {\bibinfo  {journal} {Phys. Rev. B}\ }\textbf {\bibinfo {volume}
  {102}},\ \bibinfo {pages} {104310} (\bibinfo {year} {2020})}\BibitemShut
  {NoStop}%
\bibitem [{\citenamefont {Simoncelli}\ \emph {et~al.}(2020)\citenamefont
  {Simoncelli}, \citenamefont {Marzari},\ and\ \citenamefont
  {Cepellotti}}]{simoncelli2020generalization}%
  \BibitemOpen
  \bibfield  {author} {\bibinfo {author} {\bibfnamefont {M.}~\bibnamefont
  {Simoncelli}}, \bibinfo {author} {\bibfnamefont {N.}~\bibnamefont
  {Marzari}},\ and\ \bibinfo {author} {\bibfnamefont {A.}~\bibnamefont
  {Cepellotti}},\ }\bibfield  {title} {\bibinfo {title} {{Generalization of
  Fourier's law into viscous heat equations}},\ }\href@noop {} {\bibfield
  {journal} {\bibinfo  {journal} {Phys. Rev. X}\ }\textbf {\bibinfo {volume}
  {10}},\ \bibinfo {pages} {011019} (\bibinfo {year} {2020})}\BibitemShut
  {NoStop}%
\bibitem [{\citenamefont {Srivastava}(1990)}]{srivastava2019physics}%
  \BibitemOpen
  \bibfield  {author} {\bibinfo {author} {\bibfnamefont {G.~P.}\ \bibnamefont
  {Srivastava}},\ }\href@noop {} {\emph {\bibinfo {title} {The physics of
  phonons}}}\ (\bibinfo  {publisher} {Routledge},\ \bibinfo {year}
  {1990})\BibitemShut {NoStop}%
\bibitem [{\citenamefont {Lindsay}\ \emph {et~al.}(2014)\citenamefont
  {Lindsay}, \citenamefont {Li}, \citenamefont {Carrete}, \citenamefont
  {Mingo}, \citenamefont {Broido},\ and\ \citenamefont
  {Reinecke}}]{lindsay2014phonon}%
  \BibitemOpen
  \bibfield  {author} {\bibinfo {author} {\bibfnamefont {L.}~\bibnamefont
  {Lindsay}}, \bibinfo {author} {\bibfnamefont {W.}~\bibnamefont {Li}},
  \bibinfo {author} {\bibfnamefont {J.}~\bibnamefont {Carrete}}, \bibinfo
  {author} {\bibfnamefont {N.}~\bibnamefont {Mingo}}, \bibinfo {author}
  {\bibfnamefont {D.}~\bibnamefont {Broido}},\ and\ \bibinfo {author}
  {\bibfnamefont {T.}~\bibnamefont {Reinecke}},\ }\bibfield  {title} {\bibinfo
  {title} {Phonon thermal transport in strained and unstrained graphene from
  first principles},\ }\href@noop {} {\bibfield  {journal} {\bibinfo  {journal}
  {Phys. Rev. B}\ }\textbf {\bibinfo {volume} {89}},\ \bibinfo {pages} {155426}
  (\bibinfo {year} {2014})}\BibitemShut {NoStop}%
\bibitem [{\citenamefont {Lindsay}(2016)}]{lindsay2016first}%
  \BibitemOpen
  \bibfield  {author} {\bibinfo {author} {\bibfnamefont {L.}~\bibnamefont
  {Lindsay}},\ }\bibfield  {title} {\bibinfo {title} {{First principles
  Peierls-Boltzmann phonon thermal transport: a topical review}},\ }\href@noop
  {} {\bibfield  {journal} {\bibinfo  {journal} {Nanoscale and Microscale
  Thermophysical Engineering}\ }\textbf {\bibinfo {volume} {20}},\ \bibinfo
  {pages} {67} (\bibinfo {year} {2016})}\BibitemShut {NoStop}%
\bibitem [{\citenamefont {McGaughey}\ \emph {et~al.}(2019)\citenamefont
  {McGaughey}, \citenamefont {Jain}, \citenamefont {Kim},\ and\ \citenamefont
  {Fu}}]{mcgaughey2019phonon}%
  \BibitemOpen
  \bibfield  {author} {\bibinfo {author} {\bibfnamefont {A.~J.~H.}\
  \bibnamefont {McGaughey}}, \bibinfo {author} {\bibfnamefont {A.}~\bibnamefont
  {Jain}}, \bibinfo {author} {\bibfnamefont {H.-Y.}\ \bibnamefont {Kim}},\ and\
  \bibinfo {author} {\bibfnamefont {B.}~\bibnamefont {Fu}},\ }\bibfield
  {title} {\bibinfo {title} {Phonon properties and thermal conductivity from
  first principles, lattice dynamics, and the {Boltzmann} transport equation},\
  }\href@noop {} {\bibfield  {journal} {\bibinfo  {journal} {J. App. Phys.}\
  }\textbf {\bibinfo {volume} {125}},\ \bibinfo {pages} {011101} (\bibinfo
  {year} {2019})}\BibitemShut {NoStop}%
\bibitem [{\citenamefont {Cepellotti}\ and\ \citenamefont
  {Marzari}(2016)}]{cepellotti2016thermal}%
  \BibitemOpen
  \bibfield  {author} {\bibinfo {author} {\bibfnamefont {A.}~\bibnamefont
  {Cepellotti}}\ and\ \bibinfo {author} {\bibfnamefont {N.}~\bibnamefont
  {Marzari}},\ }\bibfield  {title} {\bibinfo {title} {Thermal transport in
  crystals as a kinetic theory of relaxons},\ }\href@noop {} {\bibfield
  {journal} {\bibinfo  {journal} {Phys. Rev. X}\ }\textbf {\bibinfo {volume}
  {6}},\ \bibinfo {pages} {041013} (\bibinfo {year} {2016})}\BibitemShut
  {NoStop}%
\bibitem [{\citenamefont {Cepellotti}\ and\ \citenamefont
  {Marzari}(2017)}]{cepellotti2017transport}%
  \BibitemOpen
  \bibfield  {author} {\bibinfo {author} {\bibfnamefont {A.}~\bibnamefont
  {Cepellotti}}\ and\ \bibinfo {author} {\bibfnamefont {N.}~\bibnamefont
  {Marzari}},\ }\bibfield  {title} {\bibinfo {title} {Transport waves as
  crystal excitations},\ }\href@noop {} {\bibfield  {journal} {\bibinfo
  {journal} {Phys. Rev. Mat.}\ }\textbf {\bibinfo {volume} {1}},\ \bibinfo
  {pages} {045406} (\bibinfo {year} {2017})}\BibitemShut {NoStop}%
\bibitem [{\citenamefont {Fugallo}\ \emph {et~al.}(2013)\citenamefont
  {Fugallo}, \citenamefont {Lazzeri}, \citenamefont {Paulatto},\ and\
  \citenamefont {Mauri}}]{fugallo2013ab}%
  \BibitemOpen
  \bibfield  {author} {\bibinfo {author} {\bibfnamefont {G.}~\bibnamefont
  {Fugallo}}, \bibinfo {author} {\bibfnamefont {M.}~\bibnamefont {Lazzeri}},
  \bibinfo {author} {\bibfnamefont {L.}~\bibnamefont {Paulatto}},\ and\
  \bibinfo {author} {\bibfnamefont {F.}~\bibnamefont {Mauri}},\ }\bibfield
  {title} {\bibinfo {title} {Ab initio variational approach for evaluating
  lattice thermal conductivity},\ }\href@noop {} {\bibfield  {journal}
  {\bibinfo  {journal} {Phys. Rev. B}\ }\textbf {\bibinfo {volume} {88}},\
  \bibinfo {pages} {045430} (\bibinfo {year} {2013})}\BibitemShut {NoStop}%
\bibitem [{\citenamefont {Fugallo}\ \emph {et~al.}(2014)\citenamefont
  {Fugallo}, \citenamefont {Cepellotti}, \citenamefont {Paulatto},
  \citenamefont {Lazzeri}, \citenamefont {Marzari},\ and\ \citenamefont
  {Mauri}}]{fugallo2014thermal}%
  \BibitemOpen
  \bibfield  {author} {\bibinfo {author} {\bibfnamefont {G.}~\bibnamefont
  {Fugallo}}, \bibinfo {author} {\bibfnamefont {A.}~\bibnamefont {Cepellotti}},
  \bibinfo {author} {\bibfnamefont {L.}~\bibnamefont {Paulatto}}, \bibinfo
  {author} {\bibfnamefont {M.}~\bibnamefont {Lazzeri}}, \bibinfo {author}
  {\bibfnamefont {N.}~\bibnamefont {Marzari}},\ and\ \bibinfo {author}
  {\bibfnamefont {F.}~\bibnamefont {Mauri}},\ }\bibfield  {title} {\bibinfo
  {title} {Thermal conductivity of graphene and graphite: collective
  excitations and mean free paths},\ }\href@noop {} {\bibfield  {journal}
  {\bibinfo  {journal} {Nano letters}\ }\textbf {\bibinfo {volume} {14}},\
  \bibinfo {pages} {6109} (\bibinfo {year} {2014})}\BibitemShut {NoStop}%
\bibitem [{\citenamefont {Li}\ \emph {et~al.}(2014)\citenamefont {Li},
  \citenamefont {Carrete}, \citenamefont {Katcho},\ and\ \citenamefont
  {Mingo}}]{li2014shengbte}%
  \BibitemOpen
  \bibfield  {author} {\bibinfo {author} {\bibfnamefont {W.}~\bibnamefont
  {Li}}, \bibinfo {author} {\bibfnamefont {J.}~\bibnamefont {Carrete}},
  \bibinfo {author} {\bibfnamefont {N.~A.}\ \bibnamefont {Katcho}},\ and\
  \bibinfo {author} {\bibfnamefont {N.}~\bibnamefont {Mingo}},\ }\bibfield
  {title} {\bibinfo {title} {{ShengBTE: A solver of the Boltzmann transport
  equation for phonons}},\ }\href@noop {} {\bibfield  {journal} {\bibinfo
  {journal} {Comp. Phys. Commun.}\ }\textbf {\bibinfo {volume} {185}},\
  \bibinfo {pages} {1747} (\bibinfo {year} {2014})}\BibitemShut {NoStop}%
\bibitem [{\citenamefont {Chernatynskiy}\ and\ \citenamefont
  {Phillpot}(2015)}]{CHERNATYNSKIY2015196}%
  \BibitemOpen
  \bibfield  {author} {\bibinfo {author} {\bibfnamefont {A.}~\bibnamefont
  {Chernatynskiy}}\ and\ \bibinfo {author} {\bibfnamefont {S.~R.}\ \bibnamefont
  {Phillpot}},\ }\bibfield  {title} {\bibinfo {title} {Phonon transport
  simulator {(PhonTS)}},\ }\href@noop {} {\bibfield  {journal} {\bibinfo
  {journal} {Comp. Phys. Commun.}\ }\textbf {\bibinfo {volume} {192}},\
  \bibinfo {pages} {196} (\bibinfo {year} {2015})}\BibitemShut {NoStop}%
\bibitem [{\citenamefont {Carrete}\ \emph {et~al.}(2017)\citenamefont
  {Carrete}, \citenamefont {Vermeersch}, \citenamefont {Katre}, \citenamefont
  {van Roekeghem}, \citenamefont {Wang}, \citenamefont {Madsen},\ and\
  \citenamefont {Mingo}}]{carrete2017almabte}%
  \BibitemOpen
  \bibfield  {author} {\bibinfo {author} {\bibfnamefont {J.}~\bibnamefont
  {Carrete}}, \bibinfo {author} {\bibfnamefont {B.}~\bibnamefont {Vermeersch}},
  \bibinfo {author} {\bibfnamefont {A.}~\bibnamefont {Katre}}, \bibinfo
  {author} {\bibfnamefont {A.}~\bibnamefont {van Roekeghem}}, \bibinfo {author}
  {\bibfnamefont {T.}~\bibnamefont {Wang}}, \bibinfo {author} {\bibfnamefont
  {G.~K.~H.}\ \bibnamefont {Madsen}},\ and\ \bibinfo {author} {\bibfnamefont
  {N.}~\bibnamefont {Mingo}},\ }\bibfield  {title} {\bibinfo {title} {{almaBTE:
  A solver of the space--time dependent Boltzmann transport equation for
  phonons in structured materials}},\ }\href@noop {} {\bibfield  {journal}
  {\bibinfo  {journal} {Comp. Phys. Commun.}\ }\textbf {\bibinfo {volume}
  {220}},\ \bibinfo {pages} {351} (\bibinfo {year} {2017})}\BibitemShut
  {NoStop}%
\bibitem [{\citenamefont {Hua}\ and\ \citenamefont
  {Minnich}(2014)}]{hua2014analytical}%
  \BibitemOpen
  \bibfield  {author} {\bibinfo {author} {\bibfnamefont {C.}~\bibnamefont
  {Hua}}\ and\ \bibinfo {author} {\bibfnamefont {A.~J.}\ \bibnamefont
  {Minnich}},\ }\bibfield  {title} {\bibinfo {title} {{Analytical Green's
  function of the multidimensional frequency-dependent phonon Boltzmann
  equation}},\ }\href@noop {} {\bibfield  {journal} {\bibinfo  {journal} {Phys.
  Rev. B}\ }\textbf {\bibinfo {volume} {90}},\ \bibinfo {pages} {214306}
  (\bibinfo {year} {2014})}\BibitemShut {NoStop}%
\bibitem [{\citenamefont {Zeng}\ and\ \citenamefont
  {Chen}(2014)}]{zeng2014disparate}%
  \BibitemOpen
  \bibfield  {author} {\bibinfo {author} {\bibfnamefont {L.}~\bibnamefont
  {Zeng}}\ and\ \bibinfo {author} {\bibfnamefont {G.}~\bibnamefont {Chen}},\
  }\bibfield  {title} {\bibinfo {title} {Disparate quasiballistic heat
  conduction regimes from periodic heat sources on a substrate},\ }\href@noop
  {} {\bibfield  {journal} {\bibinfo  {journal} {J. Appl. Phys.}\ }\textbf
  {\bibinfo {volume} {116}},\ \bibinfo {pages} {064307} (\bibinfo {year}
  {2014})}\BibitemShut {NoStop}%
\bibitem [{\citenamefont {Collins}\ \emph {et~al.}(2013)\citenamefont
  {Collins}, \citenamefont {Maznev}, \citenamefont {Tian}, \citenamefont
  {Esfarjani}, \citenamefont {Nelson},\ and\ \citenamefont
  {Chen}}]{collins2013non}%
  \BibitemOpen
  \bibfield  {author} {\bibinfo {author} {\bibfnamefont {K.~C.}\ \bibnamefont
  {Collins}}, \bibinfo {author} {\bibfnamefont {A.~A.}\ \bibnamefont {Maznev}},
  \bibinfo {author} {\bibfnamefont {Z.}~\bibnamefont {Tian}}, \bibinfo {author}
  {\bibfnamefont {K.}~\bibnamefont {Esfarjani}}, \bibinfo {author}
  {\bibfnamefont {K.~A.}\ \bibnamefont {Nelson}},\ and\ \bibinfo {author}
  {\bibfnamefont {G.}~\bibnamefont {Chen}},\ }\bibfield  {title} {\bibinfo
  {title} {{Non-diffusive relaxation of a transient thermal grating analyzed
  with the Boltzmann transport equation}},\ }\href@noop {} {\bibfield
  {journal} {\bibinfo  {journal} {J. Appl. Phys.}\ }\textbf {\bibinfo {volume}
  {114}},\ \bibinfo {pages} {104302} (\bibinfo {year} {2013})}\BibitemShut
  {NoStop}%
\bibitem [{\citenamefont {Allen}(2018)}]{allen2018analysis}%
  \BibitemOpen
  \bibfield  {author} {\bibinfo {author} {\bibfnamefont {P.~B.}\ \bibnamefont
  {Allen}},\ }\bibfield  {title} {\bibinfo {title} {Analysis of nonlocal phonon
  thermal conductivity simulations showing the ballistic to diffusive
  crossover},\ }\href@noop {} {\bibfield  {journal} {\bibinfo  {journal} {Phys.
  Rev. B}\ }\textbf {\bibinfo {volume} {97}},\ \bibinfo {pages} {134307}
  (\bibinfo {year} {2018})}\BibitemShut {NoStop}%
\bibitem [{\citenamefont {Chiloyan}\ \emph {et~al.}(2021)\citenamefont
  {Chiloyan}, \citenamefont {Huberman}, \citenamefont {Ding}, \citenamefont
  {Mendoza}, \citenamefont {Maznev}, \citenamefont {Nelson},\ and\
  \citenamefont {Chen}}]{chiloyan2021green}%
  \BibitemOpen
  \bibfield  {author} {\bibinfo {author} {\bibfnamefont {V.}~\bibnamefont
  {Chiloyan}}, \bibinfo {author} {\bibfnamefont {S.}~\bibnamefont {Huberman}},
  \bibinfo {author} {\bibfnamefont {Z.}~\bibnamefont {Ding}}, \bibinfo {author}
  {\bibfnamefont {J.}~\bibnamefont {Mendoza}}, \bibinfo {author} {\bibfnamefont
  {A.~A.}\ \bibnamefont {Maznev}}, \bibinfo {author} {\bibfnamefont {K.~A.}\
  \bibnamefont {Nelson}},\ and\ \bibinfo {author} {\bibfnamefont
  {G.}~\bibnamefont {Chen}},\ }\bibfield  {title} {\bibinfo {title} {Green's
  functions of the boltzmann transport equation with the full scattering matrix
  for phonon nanoscale transport beyond the relaxation-time approximation},\
  }\href@noop {} {\bibfield  {journal} {\bibinfo  {journal} {Physical Review
  B}\ }\textbf {\bibinfo {volume} {104}},\ \bibinfo {pages} {245424} (\bibinfo
  {year} {2021})}\BibitemShut {NoStop}%
\bibitem [{\citenamefont {Balandin}\ \emph {et~al.}(2008)\citenamefont
  {Balandin}, \citenamefont {Ghosh}, \citenamefont {Bao}, \citenamefont
  {Calizo}, \citenamefont {Teweldebrhan}, \citenamefont {Miao},\ and\
  \citenamefont {Lau}}]{balandin2008superior}%
  \BibitemOpen
  \bibfield  {author} {\bibinfo {author} {\bibfnamefont {A.~A.}\ \bibnamefont
  {Balandin}}, \bibinfo {author} {\bibfnamefont {S.}~\bibnamefont {Ghosh}},
  \bibinfo {author} {\bibfnamefont {W.}~\bibnamefont {Bao}}, \bibinfo {author}
  {\bibfnamefont {I.}~\bibnamefont {Calizo}}, \bibinfo {author} {\bibfnamefont
  {D.}~\bibnamefont {Teweldebrhan}}, \bibinfo {author} {\bibfnamefont
  {F.}~\bibnamefont {Miao}},\ and\ \bibinfo {author} {\bibfnamefont {C.~N.}\
  \bibnamefont {Lau}},\ }\bibfield  {title} {\bibinfo {title} {Superior thermal
  conductivity of single-layer graphene},\ }\href@noop {} {\bibfield  {journal}
  {\bibinfo  {journal} {Nano letters}\ }\textbf {\bibinfo {volume} {8}},\
  \bibinfo {pages} {902} (\bibinfo {year} {2008})}\BibitemShut {NoStop}%
\bibitem [{\citenamefont {Chen}\ \emph {et~al.}(2011)\citenamefont {Chen},
  \citenamefont {Moore}, \citenamefont {Cai}, \citenamefont {Suk},
  \citenamefont {An}, \citenamefont {Mishra}, \citenamefont {Amos},
  \citenamefont {Magnuson}, \citenamefont {Kang}, \citenamefont {Shi} \emph
  {et~al.}}]{chen2011raman}%
  \BibitemOpen
  \bibfield  {author} {\bibinfo {author} {\bibfnamefont {S.}~\bibnamefont
  {Chen}}, \bibinfo {author} {\bibfnamefont {A.~L.}\ \bibnamefont {Moore}},
  \bibinfo {author} {\bibfnamefont {W.}~\bibnamefont {Cai}}, \bibinfo {author}
  {\bibfnamefont {J.~W.}\ \bibnamefont {Suk}}, \bibinfo {author} {\bibfnamefont
  {J.}~\bibnamefont {An}}, \bibinfo {author} {\bibfnamefont {C.}~\bibnamefont
  {Mishra}}, \bibinfo {author} {\bibfnamefont {C.}~\bibnamefont {Amos}},
  \bibinfo {author} {\bibfnamefont {C.~W.}\ \bibnamefont {Magnuson}}, \bibinfo
  {author} {\bibfnamefont {J.}~\bibnamefont {Kang}}, \bibinfo {author}
  {\bibfnamefont {L.}~\bibnamefont {Shi}}, \emph {et~al.},\ }\bibfield  {title}
  {\bibinfo {title} {Raman measurements of thermal transport in suspended
  monolayer graphene of variable sizes in vacuum and gaseous environments},\
  }\href@noop {} {\bibfield  {journal} {\bibinfo  {journal} {ACS Nano}\
  }\textbf {\bibinfo {volume} {5}},\ \bibinfo {pages} {321} (\bibinfo {year}
  {2011})}\BibitemShut {NoStop}%
\bibitem [{\citenamefont {Libbi}\ \emph {et~al.}(2020)\citenamefont {Libbi},
  \citenamefont {Bonini},\ and\ \citenamefont
  {Marzari}}]{libbi2020thermomechanical}%
  \BibitemOpen
  \bibfield  {author} {\bibinfo {author} {\bibfnamefont {F.}~\bibnamefont
  {Libbi}}, \bibinfo {author} {\bibfnamefont {N.}~\bibnamefont {Bonini}},\ and\
  \bibinfo {author} {\bibfnamefont {N.}~\bibnamefont {Marzari}},\ }\bibfield
  {title} {\bibinfo {title} {Thermomechanical properties of honeycomb lattices
  from internal-coordinates potentials: the case of graphene and hexagonal
  boron nitride},\ }\href@noop {} {\bibfield  {journal} {\bibinfo  {journal}
  {2D Materials}\ }\textbf {\bibinfo {volume} {8}},\ \bibinfo {pages} {015026}
  (\bibinfo {year} {2020})}\BibitemShut {NoStop}%
\bibitem [{\citenamefont {Bonini}\ \emph {et~al.}(2012)\citenamefont {Bonini},
  \citenamefont {Garg},\ and\ \citenamefont {Marzari}}]{Bonini2012}%
  \BibitemOpen
  \bibfield  {author} {\bibinfo {author} {\bibfnamefont {N.}~\bibnamefont
  {Bonini}}, \bibinfo {author} {\bibfnamefont {J.}~\bibnamefont {Garg}},\ and\
  \bibinfo {author} {\bibfnamefont {N.}~\bibnamefont {Marzari}},\ }\bibfield
  {title} {\bibinfo {title} {Acoustic phonon lifetimes and thermal transport in
  free-standing and strained graphene},\ }\href
  {https://doi.org/10.1021/nl202694m} {\bibfield  {journal} {\bibinfo
  {journal} {Nano Letters}\ }\textbf {\bibinfo {volume} {12}},\ \bibinfo
  {pages} {2673} (\bibinfo {year} {2012})}\BibitemShut {NoStop}%
\bibitem [{Note1()}]{Note1}%
  \BibitemOpen
  \bibinfo {note} {In the simulations, we use $N_q=120\times 120$ $q$-points to
  sample the Brillouin zone of graphene and for delta functions in Eq.~(\ref
  {Equ:FGR}) we use Gaussian broadening function $\delta (x)=\protect \qopname
  \relax o{exp}{-(x/\sigma )^2}/\protect \sqrt {\pi }\sigma $ with $\hbar
  \sigma =0.9$ meV.}\BibitemShut {Stop}%
\bibitem [{\citenamefont {Hua}\ and\ \citenamefont
  {Minnich}(2018)}]{hua2018heat}%
  \BibitemOpen
  \bibfield  {author} {\bibinfo {author} {\bibfnamefont {C.}~\bibnamefont
  {Hua}}\ and\ \bibinfo {author} {\bibfnamefont {A.~J.}\ \bibnamefont
  {Minnich}},\ }\bibfield  {title} {\bibinfo {title} {Heat dissipation in the
  quasiballistic regime studied using the boltzmann equation in the spatial
  frequency domain},\ }\href@noop {} {\bibfield  {journal} {\bibinfo  {journal}
  {Physical Review B}\ }\textbf {\bibinfo {volume} {97}},\ \bibinfo {pages}
  {014307} (\bibinfo {year} {2018})}\BibitemShut {NoStop}%
\bibitem [{\citenamefont {Chiloyan}\ \emph {et~al.}(2020)\citenamefont
  {Chiloyan}, \citenamefont {Huberman}, \citenamefont {Maznev}, \citenamefont
  {Nelson},\ and\ \citenamefont {Chen}}]{chiloyan2020thermal}%
  \BibitemOpen
  \bibfield  {author} {\bibinfo {author} {\bibfnamefont {V.}~\bibnamefont
  {Chiloyan}}, \bibinfo {author} {\bibfnamefont {S.}~\bibnamefont {Huberman}},
  \bibinfo {author} {\bibfnamefont {A.~A.}\ \bibnamefont {Maznev}}, \bibinfo
  {author} {\bibfnamefont {K.~A.}\ \bibnamefont {Nelson}},\ and\ \bibinfo
  {author} {\bibfnamefont {G.}~\bibnamefont {Chen}},\ }\bibfield  {title}
  {\bibinfo {title} {Thermal transport exceeding bulk heat conduction due to
  nonthermal micro/nanoscale phonon populations},\ }\href@noop {} {\bibfield
  {journal} {\bibinfo  {journal} {Applied Physics Letters}\ }\textbf {\bibinfo
  {volume} {116}},\ \bibinfo {pages} {163102} (\bibinfo {year}
  {2020})}\BibitemShut {NoStop}%
\bibitem [{\citenamefont {{Vermeersch}}\ and\ \citenamefont
  {{Shakouri}}(2014)}]{Vermeersch2014}%
  \BibitemOpen
  \bibfield  {author} {\bibinfo {author} {\bibfnamefont {B.}~\bibnamefont
  {{Vermeersch}}}\ and\ \bibinfo {author} {\bibfnamefont {A.}~\bibnamefont
  {{Shakouri}}},\ }\bibfield  {title} {\bibinfo {title} {{Nonlocality in
  microscale heat conduction}},\ }\href@noop {} {\bibfield  {journal} {\bibinfo
   {journal} {ArXiv e-prints}\ } (\bibinfo {year} {2014})},\ \Eprint
  {https://arxiv.org/abs/1412.6555v2} {arXiv:1412.6555v2} \BibitemShut
  {NoStop}%
\bibitem [{\citenamefont {Vermeersch}\ \emph {et~al.}(2015)\citenamefont
  {Vermeersch}, \citenamefont {Carrete}, \citenamefont {Mingo},\ and\
  \citenamefont {Shakouri}}]{VermeerschI2015}%
  \BibitemOpen
  \bibfield  {author} {\bibinfo {author} {\bibfnamefont {B.}~\bibnamefont
  {Vermeersch}}, \bibinfo {author} {\bibfnamefont {J.}~\bibnamefont {Carrete}},
  \bibinfo {author} {\bibfnamefont {N.}~\bibnamefont {Mingo}},\ and\ \bibinfo
  {author} {\bibfnamefont {A.}~\bibnamefont {Shakouri}},\ }\bibfield  {title}
  {\bibinfo {title} {Superdiffusive heat conduction in semiconductor alloys.
  {I}. {Theoretical} foundations},\ }\href@noop {} {\bibfield  {journal}
  {\bibinfo  {journal} {Phys. Rev. B}\ }\textbf {\bibinfo {volume} {91}},\
  \bibinfo {pages} {085202} (\bibinfo {year} {2015})}\BibitemShut {NoStop}%
\bibitem [{\citenamefont {Allen}\ and\ \citenamefont
  {Nghiem}(2022)}]{Allen2022}%
  \BibitemOpen
  \bibfield  {author} {\bibinfo {author} {\bibfnamefont {P.~B.}\ \bibnamefont
  {Allen}}\ and\ \bibinfo {author} {\bibfnamefont {N.~A.}\ \bibnamefont
  {Nghiem}},\ }\bibfield  {title} {\bibinfo {title} {Nonlocal phonon heat
  transport seen in 1-d pulses, to be published},\ }\href@noop {} {\bibfield
  {journal} {\bibinfo  {journal} {ArXiv e-prints}\ } (\bibinfo {year}
  {2022})},\ \Eprint {https://arxiv.org/abs/2106.00867v4} {arXiv:2106.00867v4}
  \BibitemShut {NoStop}%
\bibitem [{\citenamefont {Li}\ and\ \citenamefont
  {Lee}(2019)}]{li2019crossover}%
  \BibitemOpen
  \bibfield  {author} {\bibinfo {author} {\bibfnamefont {X.}~\bibnamefont
  {Li}}\ and\ \bibinfo {author} {\bibfnamefont {S.}~\bibnamefont {Lee}},\
  }\bibfield  {title} {\bibinfo {title} {Crossover of ballistic, hydrodynamic,
  and diffusive phonon transport in suspended graphene},\ }\href@noop {}
  {\bibfield  {journal} {\bibinfo  {journal} {Phys. Rev. B}\ }\textbf {\bibinfo
  {volume} {99}},\ \bibinfo {pages} {085202} (\bibinfo {year}
  {2019})}\BibitemShut {NoStop}%
\bibitem [{\citenamefont {Lindsay}\ and\ \citenamefont
  {Broido}(2010)}]{lindsay2010optimized}%
  \BibitemOpen
  \bibfield  {author} {\bibinfo {author} {\bibfnamefont {L.}~\bibnamefont
  {Lindsay}}\ and\ \bibinfo {author} {\bibfnamefont {D.~A.}\ \bibnamefont
  {Broido}},\ }\bibfield  {title} {\bibinfo {title} {{Optimized Tersoff and
  Brenner empirical potential parameters for lattice dynamics and phonon
  thermal transport in carbon nanotubes and graphene}},\ }\href@noop {}
  {\bibfield  {journal} {\bibinfo  {journal} {Phys. Rev. B}\ }\textbf {\bibinfo
  {volume} {81}},\ \bibinfo {pages} {205441} (\bibinfo {year}
  {2010})}\BibitemShut {NoStop}%
\bibitem [{\citenamefont {Gu}\ \emph {et~al.}(2019)\citenamefont {Gu},
  \citenamefont {Fan}, \citenamefont {Bao},\ and\ \citenamefont
  {Zhao}}]{Gu2019}%
  \BibitemOpen
  \bibfield  {author} {\bibinfo {author} {\bibfnamefont {X.}~\bibnamefont
  {Gu}}, \bibinfo {author} {\bibfnamefont {Z.}~\bibnamefont {Fan}}, \bibinfo
  {author} {\bibfnamefont {H.}~\bibnamefont {Bao}},\ and\ \bibinfo {author}
  {\bibfnamefont {C.~Y.}\ \bibnamefont {Zhao}},\ }\bibfield  {title} {\bibinfo
  {title} {Revisiting phonon-phonon scattering in single-layer graphene},\
  }\href {https://doi.org/10.1103/PhysRevB.100.064306} {\bibfield  {journal}
  {\bibinfo  {journal} {Phys. Rev. B}\ }\textbf {\bibinfo {volume} {100}},\
  \bibinfo {pages} {064306} (\bibinfo {year} {2019})}\BibitemShut {NoStop}%
\bibitem [{\citenamefont {Perebeinos}\ and\ \citenamefont
  {Tersoff}(2009)}]{Valencemodel2009}%
  \BibitemOpen
  \bibfield  {author} {\bibinfo {author} {\bibfnamefont {V.}~\bibnamefont
  {Perebeinos}}\ and\ \bibinfo {author} {\bibfnamefont {J.}~\bibnamefont
  {Tersoff}},\ }\bibfield  {title} {\bibinfo {title} {Valence force model for
  phonons in graphene and carbon nanotubes},\ }\href
  {https://doi.org/10.1103/PhysRevB.79.241409} {\bibfield  {journal} {\bibinfo
  {journal} {Phys. Rev. B}\ }\textbf {\bibinfo {volume} {79}},\ \bibinfo
  {pages} {241409} (\bibinfo {year} {2009})}\BibitemShut {NoStop}%
\bibitem [{\citenamefont {Johnson}\ \emph {et~al.}(2013)\citenamefont
  {Johnson}, \citenamefont {Maznev}, \citenamefont {Cuffe}, \citenamefont
  {Eliason}, \citenamefont {Minnich}, \citenamefont {Kehoe}, \citenamefont
  {Torres}, \citenamefont {Chen},\ and\ \citenamefont
  {Nelson}}]{johnson2013direct}%
  \BibitemOpen
  \bibfield  {author} {\bibinfo {author} {\bibfnamefont {J.~A.}\ \bibnamefont
  {Johnson}}, \bibinfo {author} {\bibfnamefont {A.~A.}\ \bibnamefont {Maznev}},
  \bibinfo {author} {\bibfnamefont {J.}~\bibnamefont {Cuffe}}, \bibinfo
  {author} {\bibfnamefont {J.~K.}\ \bibnamefont {Eliason}}, \bibinfo {author}
  {\bibfnamefont {A.~J.}\ \bibnamefont {Minnich}}, \bibinfo {author}
  {\bibfnamefont {T.}~\bibnamefont {Kehoe}}, \bibinfo {author} {\bibfnamefont
  {C.~M.~S.}\ \bibnamefont {Torres}}, \bibinfo {author} {\bibfnamefont
  {G.}~\bibnamefont {Chen}},\ and\ \bibinfo {author} {\bibfnamefont {K.~A.}\
  \bibnamefont {Nelson}},\ }\bibfield  {title} {\bibinfo {title} {Direct
  measurement of room-temperature nondiffusive thermal transport over micron
  distances in a silicon membrane},\ }\href@noop {} {\bibfield  {journal}
  {\bibinfo  {journal} {Phys. Rev. Lett.}\ }\textbf {\bibinfo {volume} {110}},\
  \bibinfo {pages} {025901} (\bibinfo {year} {2013})}\BibitemShut {NoStop}%
\bibitem [{\citenamefont {Zeng}\ \emph {et~al.}(2015)\citenamefont {Zeng},
  \citenamefont {Collins}, \citenamefont {Hu}, \citenamefont {Luckyanova},
  \citenamefont {Maznev}, \citenamefont {Huberman}, \citenamefont {Chiloyan},
  \citenamefont {Zhou}, \citenamefont {Huang}, \citenamefont {Nelson} \emph
  {et~al.}}]{zeng2015measuring}%
  \BibitemOpen
  \bibfield  {author} {\bibinfo {author} {\bibfnamefont {L.}~\bibnamefont
  {Zeng}}, \bibinfo {author} {\bibfnamefont {K.~C.}\ \bibnamefont {Collins}},
  \bibinfo {author} {\bibfnamefont {Y.}~\bibnamefont {Hu}}, \bibinfo {author}
  {\bibfnamefont {M.~N.}\ \bibnamefont {Luckyanova}}, \bibinfo {author}
  {\bibfnamefont {A.~A.}\ \bibnamefont {Maznev}}, \bibinfo {author}
  {\bibfnamefont {S.}~\bibnamefont {Huberman}}, \bibinfo {author}
  {\bibfnamefont {V.}~\bibnamefont {Chiloyan}}, \bibinfo {author}
  {\bibfnamefont {J.}~\bibnamefont {Zhou}}, \bibinfo {author} {\bibfnamefont
  {X.}~\bibnamefont {Huang}}, \bibinfo {author} {\bibfnamefont {K.~A.}\
  \bibnamefont {Nelson}}, \emph {et~al.},\ }\bibfield  {title} {\bibinfo
  {title} {Measuring phonon mean free path distributions by probing
  quasiballistic phonon transport in grating nanostructures},\ }\href@noop {}
  {\bibfield  {journal} {\bibinfo  {journal} {Sci. Rep.}\ }\textbf {\bibinfo
  {volume} {5}},\ \bibinfo {pages} {1} (\bibinfo {year} {2015})}\BibitemShut
  {NoStop}%
\bibitem [{\citenamefont {Li}\ \emph {et~al.}(2019)\citenamefont {Li},
  \citenamefont {Xiong}, \citenamefont {Sievers}, \citenamefont {Hu},
  \citenamefont {Fan}, \citenamefont {Wei}, \citenamefont {Bao}, \citenamefont
  {Chen}, \citenamefont {Donadio},\ and\ \citenamefont
  {Ala-Nissila}}]{li2019influence}%
  \BibitemOpen
  \bibfield  {author} {\bibinfo {author} {\bibfnamefont {Z.}~\bibnamefont
  {Li}}, \bibinfo {author} {\bibfnamefont {S.}~\bibnamefont {Xiong}}, \bibinfo
  {author} {\bibfnamefont {C.}~\bibnamefont {Sievers}}, \bibinfo {author}
  {\bibfnamefont {Y.}~\bibnamefont {Hu}}, \bibinfo {author} {\bibfnamefont
  {Z.}~\bibnamefont {Fan}}, \bibinfo {author} {\bibfnamefont {N.}~\bibnamefont
  {Wei}}, \bibinfo {author} {\bibfnamefont {H.}~\bibnamefont {Bao}}, \bibinfo
  {author} {\bibfnamefont {S.}~\bibnamefont {Chen}}, \bibinfo {author}
  {\bibfnamefont {D.}~\bibnamefont {Donadio}},\ and\ \bibinfo {author}
  {\bibfnamefont {T.}~\bibnamefont {Ala-Nissila}},\ }\bibfield  {title}
  {\bibinfo {title} {Influence of thermostatting on nonequilibrium molecular
  dynamics simulations of heat conduction in solids},\ }\href@noop {}
  {\bibfield  {journal} {\bibinfo  {journal} {The Journal of chemical physics}\
  }\textbf {\bibinfo {volume} {151}},\ \bibinfo {pages} {234105} (\bibinfo
  {year} {2019})}\BibitemShut {NoStop}%
\bibitem [{\citenamefont {Cahill}(2004)}]{Cahill2004}%
  \BibitemOpen
  \bibfield  {author} {\bibinfo {author} {\bibfnamefont {D.~G.}\ \bibnamefont
  {Cahill}},\ }\bibfield  {title} {\bibinfo {title} {Analysis of heat flow in
  layered structures for time-domain thermoreflectance},\ }\href@noop {}
  {\bibfield  {journal} {\bibinfo  {journal} {Rev. Sci. Instr.}\ }\textbf
  {\bibinfo {volume} {75}},\ \bibinfo {pages} {5119} (\bibinfo {year}
  {2004})}\BibitemShut {NoStop}%
\bibitem [{\citenamefont {Hoogeboom-Pot}\ \emph {et~al.}(2015)\citenamefont
  {Hoogeboom-Pot}, \citenamefont {Hernandez-Charpak}, \citenamefont {Gu},
  \citenamefont {Frazer}, \citenamefont {Anderson}, \citenamefont {Chao},
  \citenamefont {Falcone}, \citenamefont {Yang}, \citenamefont {Murnane},
  \citenamefont {Kapteyn} \emph {et~al.}}]{hoogeboom2015new}%
  \BibitemOpen
  \bibfield  {author} {\bibinfo {author} {\bibfnamefont {K.~M.}\ \bibnamefont
  {Hoogeboom-Pot}}, \bibinfo {author} {\bibfnamefont {J.~N.}\ \bibnamefont
  {Hernandez-Charpak}}, \bibinfo {author} {\bibfnamefont {X.}~\bibnamefont
  {Gu}}, \bibinfo {author} {\bibfnamefont {T.~D.}\ \bibnamefont {Frazer}},
  \bibinfo {author} {\bibfnamefont {E.~H.}\ \bibnamefont {Anderson}}, \bibinfo
  {author} {\bibfnamefont {W.}~\bibnamefont {Chao}}, \bibinfo {author}
  {\bibfnamefont {R.~W.}\ \bibnamefont {Falcone}}, \bibinfo {author}
  {\bibfnamefont {R.}~\bibnamefont {Yang}}, \bibinfo {author} {\bibfnamefont
  {M.~M.}\ \bibnamefont {Murnane}}, \bibinfo {author} {\bibfnamefont {H.~C.}\
  \bibnamefont {Kapteyn}}, \emph {et~al.},\ }\bibfield  {title} {\bibinfo
  {title} {A new regime of nanoscale thermal transport: Collective diffusion
  increases dissipation efficiency},\ }\href@noop {} {\bibfield  {journal}
  {\bibinfo  {journal} {Proc. Nat. Acad. Sci.}\ }\textbf {\bibinfo {volume}
  {112}},\ \bibinfo {pages} {4846} (\bibinfo {year} {2015})}\BibitemShut
  {NoStop}%
\bibitem [{\citenamefont {Lundstrom}(2000)}]{lundstrom_2000}%
  \BibitemOpen
  \bibfield  {author} {\bibinfo {author} {\bibfnamefont {M.}~\bibnamefont
  {Lundstrom}},\ }\href@noop {} {\emph {\bibinfo {title} {Fundamentals of
  Carrier Transport}}},\ \bibinfo {edition} {2nd}\ ed.\ (\bibinfo  {publisher}
  {Cambridge University Press},\ \bibinfo {year} {2000})\BibitemShut {NoStop}%
\bibitem [{\citenamefont {{Center for Computational Research, University at
  Buffalo}}()}]{UBCCR}%
  \BibitemOpen
  \bibfield  {author} {\bibinfo {author} {\bibnamefont {{Center for
  Computational Research, University at Buffalo}}},\ }\href@noop {} {}\bibinfo
  {note} {\url{http://hdl.handle.net/10477/79221}}\BibitemShut {NoStop}%
\end{thebibliography}%

\end{document}